\documentclass[aoas,MSNbibl,nameyear,seceqn,dvips]{arximspdf}
\usepackage{graphicx}

\doi{10.1214/12-AOAS565} 
\volume{6}
\issue{4}
\pubyear{2012}
\firstpage{1906}
\lastpage{1948}

\makeatletter

\newcommand{\iint}{\int\!\!\int}

\newcommand{\rrVert}{\Vert}
\newcommand{\rrvert}{\vert}
\newcommand{\llVert}{\Vert}
\newcommand{\llvert}{\vert}

\newcommand{\implies}{\Longrightarrow}

\newcommand{\tr}{\operatorname{tr}}

\newcommand{\E}{\mathrm{E}}
\newcommand{\cov}{\operatorname{cov}}

\newcommand{\dto}{\stackrel{\mathcal{L}}{\longrightarrow}}
\newcommand{\pto}{\stackrel{P}{\longrightarrow}}

\newcommand{\vth}{\vartheta}

\newtheorem{theorem}{Theorem}[section]
\newtheorem{cor}{Corollary}[section]

\newproclaim{rem}{Remark}[section]

\makeatother

\begin{document}
\begin{frontmatter}

\title{Evaluating stationarity via change-point alternatives with
applications to fMRI data}
\runtitle{Stationarity, change points and fMRI}

\begin{aug}
\author[A]{\fnms{John A. D.} \snm{Aston}\corref{}\thanksref{T1}\ead[label=e1]{j.a.d.aston@warwick.ac.uk}}
\and
\author[B]{\fnms{Claudia} \snm{Kirch}\thanksref{T2}\ead[label=e2]{claudia.kirch@kit.edu}}
\runauthor{J. A. D. Aston and C. Kirch}
\affiliation{University of Warwick and Karlsruhe Institute of Technology}
\address[A]{CRiSM\\
Department of Statistics\\
University of Warwick\\
Coventry\\
CV4 7AL\\
United Kingdom\\
\printead{e1}}
\address[B]{Institute for Stochastics\\
Karlsruhe Institute of Technology (KIT)\\
Kaiserstr. 89\\
D-76133 Karlsruhe\\
Germany\\
\printead{e2}} 
\end{aug}

\thankstext{T1}{Supported by the
Engineering and Physical Sciences Research Council (UK) through the
CRiSM programme grant and by the project Grant EP/H016856/1.}

\thankstext{T2}{Supported by the Stifterverband f\"ur die Deutsche
Wissenschaft by funds of the Claussen-Simon-trust.}

\received{\smonth{9} \syear{2011}}
\revised{\smonth{4} \syear{2012}}

%
\begin{abstract}
Functional magnetic resonance imaging (fMRI) is now a well-established
technique for studying the brain. However, in many situations, such as
when data are acquired in a resting state, it is difficult to know
whether the data are truly stationary or if level shifts have occurred.
To this end, change-point detection in sequences of functional data is
examined where the functional observations are dependent and where the
distributions of change-points from multiple subjects are required. Of
particular interest is the case where the change-point is an epidemic
change---a change occurs and then the observations return to baseline
at a later time. The case where the covariance can be decomposed as a
tensor product is considered with particular attention to the power
analysis for detection. This is of interest in the application to fMRI,
where the estimation of a full covariance structure for the
three-dimensional image is not computationally feasible. Using the
developed methods, a large study of resting state fMRI data is
conducted to determine whether the subjects undertaking the resting
scan have nonstationarities present in their time courses. It is found
that a sizeable proportion of the subjects studied are not stationary.
The change-point distribution for those subjects is empirically
determined, as well as its theoretical properties examined.
\end{abstract}

%

\end{frontmatter}

\section{Introduction}

An increasing number of applications from biology to image sequences in
medical imaging involve data that can be well represented as functional
time series. This has led to a rapid progression of theory associated
with functional data, particularly regarding complex correlation
structures present within and across many observed functional data.
These structures require methods that can deal both with internal and
external dependencies between the observations. Nonparametric
techniques for the analysis of functional data are becoming well
established [see \citet{FerratyV2006} or \citet{horkokbook}
for a good overview], and this paper sets out a nonparametric framework
for change-point analysis within and across dependent functional data.

Given its generality, applications for the methodology are fairly
widespread, but in this paper, we are, in particular, interested in
functional magnetic resonance imaging (fMRI), an image acquisition
modality used to study the brain in-vivo. fMRI is concerned with
characterizing relative blood flow changes, based on the changes in the
proportions of oxy- and deoxy-hemoglo\-bin levels in regions of the
brain, the so-called Blood Oxygenation Level Dependent (BOLD) response
[\citet{Ogawa90}]. Changes in the BOLD response can be used as a
surrogate indirect measure of brain (neuronal) activity due to the
increased need for oxygen being associated with neuronal activation.
Change-point analysis has recently been highlighted as a useful
technique in fMRI [\citet{LindquistWW2007}, \citet{RobinsonWL2010}] where
different subjects react differently to stimuli such as stress or
anxiety (as the time of brain state change is much less clearly linked
to the stimuli than in an experiment involving movement, e.g.,
where the observed movement and brain activity will be intrinsically
linked). A particular type of experiment that has recently become very
popular is the resting state scan, where subjects are imaged while
lying in the scanner ``at rest.'' These data are used to infer
connections in the brain which are not due to external stimuli; see,
for example, \citet{DamoiseauxRBSSSB2006}. This amounts statistically to
an investigation of covariance structures between brain regions, which
heavily relies on the brain activity being stationary. In this paper,
we establish a framework for testing whether this is the case or
whether the observed time series contain level shifts, including
segments which return to the original state after some unspecified
duration. The latter activation-baseline pattern is a standard
assumption in most fMRI experiments.\looseness=-1

Time series obtained in fMRI studies typically contain all the features
with which functional data analysis is concerned. The data are
autocorrelated, recorded at a large number of locations with the
associated spatial dependencies, where these spatial data are
intrinsically discretized records of a functional response (the brain
as a whole). Modeling the brain as a single (albeit very complex three
dimensional) function is a natural representation, as the brain works
as a single unit rather than a disconnected series of voxels [voxel
(volume element)---3D element within an image, similar to a pixel in a
2D picture]. While the ``functional'' in ``f'' MRI refers to time, in
all the descriptions in this paper, the functional data is the whole
brain as a three-dimensional object, while the observations at different
time points are referred to as the time series.\vadjust{\goodbreak}

In most activation fMRI studies, responses are modeled using linear
regression and a known experimental design matrix, but in some cases,
such as those with resting state data, no experimental design is known.
Indeed, in such situations, the hypothesis of whether the data are
stationary is of interest, in that subsequent analyses often involve
empirical covariances which make little sense in the presence of
nonstationarities. Since level shifts and, in particular, epidemic
changes in the mean are a reasonable alternative to stationarity as a
first approximation for fMRI, change-point techniques become
increasingly relevant, with a need to extend the analysis to cases
beyond at most one change (AMOC). However, most change-point techniques
are not particularly designed for functional data. A considerable
amount of literature deals with process control using change-point
techniques starting as early as \citet{Page1954}. Most of these
methodologies are based on an assumed underlying model (such as i.i.d.
errors or autocorrelated error structures, e.g.) for univariate or
multivariate time series. While in many applications, the error
structure is well known, in fMRI there is still considerable
controversy where everything from AR($p$) errors
[\citet{Worsleyetal2002}] to fractional noise error processes
[\citet{Bullmoreetal2003}] have been proposed. Unlike in classic
process control techniques, in the present paper we do not assume a
specific parametric error structure but revert to nonparametric weak
dependent errors in order to limit the assumptions made. In addition,
if univariate tests are considered at each voxel location in the brain,
the important issue of multiple comparisons requires attention. By
contrast, when assuming functional observations, the brain is treated
as a whole, thus circumventing this problem. Epidemics is another area
where considerable use of change-point theory has been made. In this
context, change-point detection is usually based on the theory of
Poisson point processes [see, e.g., \citet{Diggleetal2005}], which
has distinct advantages when the data are sparsely and irregularly
sampled in both time and space, with a small number of possible spatial
locations for changes. However, in fMRI, the data are very densely
sampled and changes could take place on either a small or large spatial
scale, making such Poisson models more difficult to specify.

Current change-point methodology for fMRI data is applied voxelwise
across spatial locations to find epidemic changes using process control
theory [\citet{RobinsonWL2010}], requiring a mass univariate approach
for this very high-dimensional multivariate or functional data, with
all the problems that then ensue (particularly of spatially correlated
multiple comparisons and having to choose an error structure). For this
reason, the nonparametric functional approach considered here is of
particular interest in the analysis of fMRI data. By considering each
complete image (approximately $10^5$ observations) as a single
functional observation, we derive a true functional change detection
procedure under a weak dependent error process model.\vadjust{\goodbreak} However, to
achieve this computationally, it is necessary to incorporate the
three-dimensional spatial structure of the observations to estimate the
covariance functions required. This motivates our investigation of the
multidimensional separable structures derived in this paper.

The paper comprises three main ideas, each of which alone provides
methodology with application to fMRI analysis, and combined enable a
complete estimation of the distribution of the time of structural
breaks across a number of fMRI subjects.

First, in Section~\ref{secprojs} orthonormal projections for functional
data are investigated. Tensor based separable covariance functions for
image data are developed, giving rise to separable projections. Tensor
based methods have been previously considered in neuroimaging data
[\citet{AstonGunn2005}, \citet{BeckmannSmith05}], but not
where the tensor products are taken over functions rather than vector
spaces. Indeed, the use of functional data representations of the
entire brain is not a particularly well studied idea, with it only
being explicitly considered in a few papers, as, for example, in
\citet{Zipunnikovetal2010}.

The second idea, given in Section~\ref{secmodel}, is that of using
change-point analysis for functional data within fMRI. Epidemic
change-points are shown to be a good starting point as an alternative
to stationarity in fMRI and the resulting theory integrating separable
projections and epidemic changes provides considerable insight into the
performance of the estimators in practice. While the use of separable
projections would not be limited to change-point analysis, it is shown
here that they have particularly appropriate properties in this case,
in that a large enough separable change will switch the estimated
system in such a way that the change is no longer orthogonal to the
projection subspace making the change detectable (cf. Corollary~\ref
{lemseppower}). However, due to the small number of time observations
relative to the number of brain location observations, small sample
properties of the tests and estimators are investigated in Section \ref
{sectionbootstrap} and a revised, more robust, change-point test
introduced to alleviate estimation issues. The preceding analysis all
takes place for a single subject.

The final idea, expanded in Section~\ref{sectionhierarchical}, allows
the combination of multiple subjects' change-point times, to evaluate a
distribution of the change-point times across the population of
subjects. In many applications, such as fMRI, sets of functional
observations are recorded from a number of subjects indicating a
hierarchical structure, and the distribution of the change-points over
all subjects is an item of interest [\citet{RobinsonWL2010}]. In
addition to giving consistent estimators within one set of dependent
observations, in Section~\ref{sectiondensity} those estimators are
used to find the distribution as well as density of the change-points
in hierarchical models, where several independent sets of time series
including a random change are observed. In this case empirical
distribution functions and kernel density estimators based on the
estimated change-points for each individual time series yield
consistent results (cf. Theorems~\ref{lemdfest} and~\ref{lemdensest}).\vadjust{\goodbreak}

The data analysis of nearly 200 resting state scans is given throughout
the paper as the methodology is developed. In Section \ref
{secsinglefmri} details about the data set are given and examples of
data shown. In Section~\ref{subsectionsepconnect} examples indicate
that epidemic changes are indeed a good first approximation to the
deviation from stationarity that can be expected. Even though the scans
are not sparsely represented in terms of basis functions, only a very
small number of basis functions are needed to detect change-points in
practice (which confirms our theoretic results). In Section~\ref
{secfMRIboot} the test results for the data are reported indicating
that 40--50\% of the resting scans exhibit deviations from
stationarity, even after correction for multiple comparisons across
subjects. This indicates that substantial care should be taken when
combining resting state scans, as nonstationarities will likely be
present and these could greatly confound analyses based on
correlations, for example. Finally, in Section~\ref{sectionfMRI} the
estimators for the position and duration of the change are given for
those data sets that contained evidence of an epidemic change showing
various patterns of locations and durations for the change-points in
the 200 subject sample.

For most sections, the amount of mathematical detail has been kept
somewhat minimal to hopefully make the material more accessible.
However, Section~\ref{secstatsprops} explains theoretical details
behind the statistical ideas in this paper, justifying our proposed
analysis for fMRI data. These insights explain why only a tiny fraction
of the data's variance is used for the change-point procedure. Should
the implementation of the procedure for fMRI be most of interest, then
this section could be skipped on first reading. However, to the reader
who is interested in applying the procedure in different applications,
this section is likely to be essential to determine whether the
assumptions required are justified in another application. In addition
to the main paper, the electronic supplementary material
[\citet{AKsupp}] contains some further information regarding the
more technical details of the estimation procedures as well as the
proofs of the results in the paper.

\section{Functional magnetic resonance imaging: 1000 connectome resting
state data}\label{secsinglefmri}

To obtain a resting state scan, an individual is asked to lie in the
scanner for a period of time, usually with their eyes closed, and asked
to think of nothing in particular while not falling asleep [see, e.g.,
\citet{DamoiseauxRBSSSB2006}]. Scans of this type are used to
study the brain regions that are involved in the underlying brain
activity, also sometimes known as the default network. Various
techniques used to determine this network either explicitly or
implicitly rely on stationarity of the time series [see \citet
{ColeSB2010} for an overview of the current methods of analysis and
pitfalls associated with them]. However, it is not known whether the
areas just exhibit some stationary variation, or whether there are
changes in activity during the scan that are more than could be
expected just as a result of stationary variability. Indeed, it has
been recently postulated that the resting state network itself might be
nonstationary, with different modes of the network active dependent on
the thought processes at the time [see work by \citet{Doucetetal2012}
and \citet{Vanhaudenhuyse2010} for examples of changes in activation
patterns during resting state scans].

Consequently, stationarity in the time series can be seen as a crucial
assumption for this kind of analysis but is by no means guaranteed.
Imagine, for example, that a strong stimulus affected the subject while
undergoing the scan, such as a loud unexpected noise occurring during
the scanning session or the person suddenly recollecting they had
forgotten something important. In such cases, the activation level of
those regions processing these stimuli will change at the same time,
falsely indicating a strong correlation between these regions in a
resting state, which is in no way linked to the default network.
However, even in the default network, there is evidence that switches
take place when the mind starts wandering [\citet{Doucetetal2012}]. The
thought processes of people in the scanner are thus unlikely to always
be stationary, and, as such, tests to determine possible positions of
nonstationarity would enable these changes to be taken into account.

We use data from the 1000 connectome project which are publicly
available\setcounter{footnote}{2}\footnote{The data can be accessed at
\url{http://www.nitrc.org/projects/fcon\_1000/}.} [\citet{Biswal2010}]. This
project consists of in excess of 1200 resting state data sets. However,
a subset of this data will be used here so that confounding factors
such as different scanner types and different locations of the subjects
can be ignored. The data used were from a single site (Beijing, China)
and consist of 198 resting state scans, each comprising 225 time points
of a three-dimensional image of size $64 \times64\times33$ voxels with
each temporal scan being taken 2 seconds apart (1 scan was discarded
due to a different orientation of reconstruction, leaving 197 scans in
the analysis below).
Each scan had a polynomial trend of order 3 removed from each voxel
time series prior to estimation to remove scanner drift and other low
frequency components [\citet{Worsleyetal2002}], in addition to being
corrected for motion using the FSL software library [\citet{FSL}].

\begin{figure}
\begin{tabular}{@{}c@{}}

\includegraphics{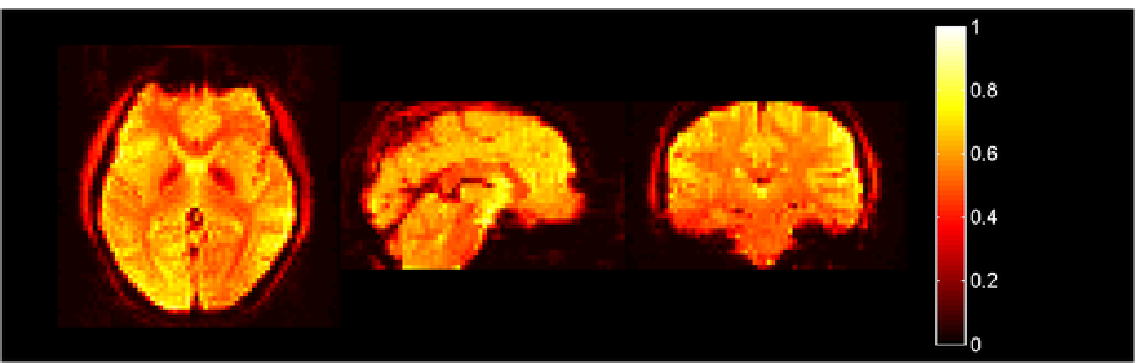}
\\
\textup{(a)}\\[6pt]

\includegraphics{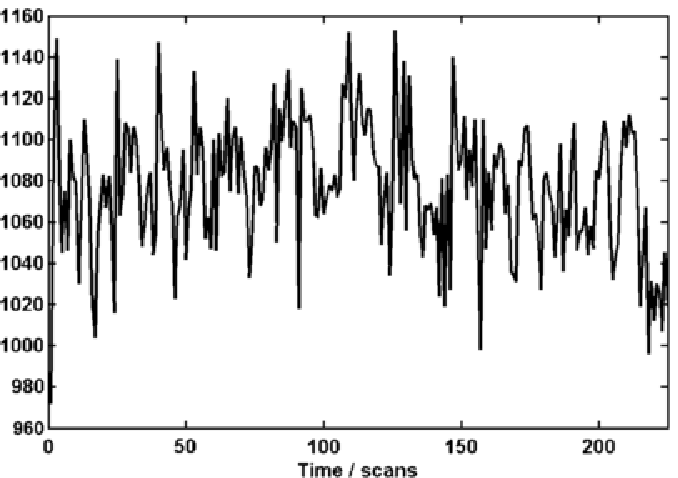}
\\
\textup{(b)}
\end{tabular}
\caption{Example of fMRI data set for a single subject from the 1000
connectome Beijing data. The top image \textup{(a)} shows a
three-dimensional view
of the temporally averaged brain data and is formally equivalent to the
mean function for these data. The bottom image \textup{(b)} shows the
time series
for the central voxel of the image. The analysis considers all the time
series of all the voxels together (as a single functional time
series).}\label{fdata}
\end{figure}

In Figure~\ref{fdata} an example of the connectome data can be seen.
The data set is a four-dimensional volume, with three spatial
dimensions and one temporal dimension. At each spatial location, there
is a recording of a time series, or, more relevantly for our functional
data analysis, for each time, there is a complete three-dimensional
%
volume present. In this paper, we will consider the spatial data as a
function, and the time series to be repeated (and correlated)
observations of that function. This implies that the spatial covariance
function will be six-dimensional, and it is this covariance that is
intrinsically of interest in resting state fMRI studies (as
connectivity maps are simply approximations to this covariance).\vadjust{\goodbreak} While
it might be possible in individual cases to use a supercomputer to
handle matrices of this order [see \citet{Longetal2011}, e.g.],
in most cases where there are large numbers of subjects to process, an
approximation, or, equivalently, dimension reduction, will be needed,
and this will now be the focus of the next section.

\section{Projections for functional data}\label{secprojs}

In functional neuroimaging, two main analysis options are usually
considered: mass univariate analysis or projection subspace analysis.
Examples of the second include analyses such as those using eigenimages
[\citet{Fristonetal1993}] and independent component analysis (ICA) [see
\citet{BeckmannSmith05}, e.g.]. In this paper, a projection
subspace approach will be taken.

In this section we detail some projections that can be used for
functional observations $X_t(u),u\in\mathcal{U}$, $t=1,\ldots,n$, where
$\mathcal{U}$ is some compact set. In fMRI, this corresponds to $u$
being the voxel location in the brain (or a continuous analogue of a
voxel), while $t$ is the scan number from the total $n$ scans taken.
Thus, the complete brain itself is treated as a single function and
this function is observed $n$ times. Of course, due to slice timing
events and voxel discretizations, this will be an approximation, but
one which naturally encodes the brain as a single observed unit.
However, should a high-dimensional multivariate approach be preferred,
the results in this paper will equally apply. In addition, we will
assume that each fMRI observation is made up of a common mean function
$\mu(u)$, that is, $X_t(u) = \mu(u) + Y_t(u)$ where $Y_t(u)$ are
deviations from the mean [with assumed mathematical properties $\{
Y_t(u)\dvtx 1\leq t\leq n\}$ are elements of $L^2(\mathcal{U})$, $\E Y_t(u)=0$
and form a stationary time series].

Below, we will define an orthonormal system $\{\widehat{v}_j(\cdot
),j=1,\ldots,d\}$ for the projection components. Associated with each
system is the score, which is determined by the inner product of the
data with the component, $\widehat{\eta}_{t,l}:=\langle X_t, \widehat
{v}_l\rangle=\int_{\mathcal{U}}X_t(u)\widehat{v}_l(u)\,du$.

The orthonormal system could be either chosen in advance, as in, for
example, wavelet based methods for functional data, or derived from the
data, as in functional principal components. In particular, if a region
based analysis in fMRI of connectivity was of interest, then the
regions of interest can be expressed as a projection of the original
data. In such a situation, whether this regional data is stationary or
not is the key question, and thus the tests of the next section should
be applied using this projection. Otherwise, if the stationarity of the
complete data is of interest rather than that of a specific projection,
then a projection should be chosen that also contains the
nonstationarities. Possible methods for choosing bases include
principal component analysis (PCA) or ICA. ICA is very popular in
resting state analyses [\citet{BeckmannDDS2005}], but PCA is often a
preprocessing step in the ICA analysis and additionally is very much
linked to the analysis of covariances, which plays a prime role in
connectivity analysis, and therefore we shall concentrate on PCA here.
It will also be shown in Section~\ref{secpowsep} that estimating the
projections using PCA can have good power for detecting
nonstationarites. As estimation of PCA components is more complex than
nonestimated bases, we will concentrate on this case (with analogous
results for the testing and estimation procedures of Section \ref
{secmodel} following in the nonestimated basis function case).

\subsection{Principal components}

Classical dimension reduction techniques are often based on the first
$d$ principal components, which choose a subspace explaining most
variance for any subspace of an equivalent dimension. The notation
below is in terms of integrals, which is simply the function based
analogue of traditional multivariate vector based PCA. To elaborate,
consider the (spatial) covariance kernel of $Y_t(\cdot)$ given by
%
\begin{equation}
\label{eqcovfunct} c(u,s)=\E\bigl(Y_t(u) Y_t(s)\bigr)
\end{equation}
and define the covariance operator $C\dvtx\mathcal{L}^2(\mathcal{U})\to
\mathcal{L}^2(\mathcal{U})$ by $C z=\int_{\mathcal{U}}c(\cdot
,s)z(s) \,ds$.\vadjust{\goodbreak}

Let $\{\lambda_k\}$ be the nonnegative decreasing sequence of
eigenvalues of the covariance operator and $\{v_k(\cdot)\dvtx k\geq1\}$ a
given set of corresponding orthonormal eigenfunctions, that is,
%
\begin{equation}
\label{eqdefeigen} \int c(u,s)v_l(s) \,ds=\lambda_lv_l(u),\qquad
l=1,2,\ldots, u\in \mathcal{U}.
\end{equation}

$Y_t(\cdot)$ can be expressed in terms of the eigenfunctions
%
\begin{equation}
\label{eqkarhloeve} Y_t(u)=\sum_{l=1}^{\infty}
\eta_{t,l} v_l(u),
\end{equation}
where 
%
\begin{equation}
\label{eqkarhloeve2} \eta_{t,l}=\int Y_t(u)v_l(u)
\,du,\qquad t=1, \ldots,n, l=1,2,\ldots,
\end{equation}
are uncorrelated with mean $0$ and variance $\lambda_l$. More details
can, for example, be found in either \citet{bosq2000} or \citet{horkokbook}.

A natural estimator in a general nonparametric setting is the empirical
version of the covariance function (analogously to standard PCA)
%
\begin{equation}
\label{eqcovnonparam} \widehat{c}_n(u,s)=\frac1 n \sum
_{t=1}^n\bigl(X_t(u)-\bar
{X}_n(u)\bigr) \bigl(X_t(s)-\bar{X}_n(s)
\bigr),
\end{equation}
where
$\bar{X}_n(u)=\frac1 n \sum_{t=1}^nX_t(u)$. 

Usually one converts the continuous functional eigenanalysis problem to
an approximately equivalent matrix eigenanalysis task. The simplest
solution is a discretization of the observed function on a fine grid.
Many data sets in applications are already obtained in this way, as in
the example of fMRI data used in this paper. For a discussion of this
as well as more advanced options, we refer to \citet{ramsaysilverman}.
In such examples of very high-dimensional data, a PCA based on the
empirical covariance matrix is computationally infeasible due to the
even higher-dimensionality of the covariance matrix. The following
computational trick can be applied but also shows the limitations of
the approach, as the number of nonzero eigenvalues of the estimated
covariance matrix is limited by the sample size, with the associated
problems for small sample sizes.

Assume that after discretization the data are given by
$X_t:=(X_t(1),\ldots,\allowbreak X_t(M))^T$, $t=1,\ldots,n$. In fMRI, $M$ here
would be the total number of voxels. The eigenanalysis problem
corresponding to the estimated covariance kernel in (\ref
{eqcovnonparam}) is to find the eigenvalues of the $M\times M$-matrix
$ZZ^T$, where $Z=(X_1-\bar{X}_n,\ldots,X_n-\bar{X}_n)$ is a $M\times
n$-matrix. One can check that $ZZ^T$ has $\operatorname{rank}(Z)\leq\min(M,n)$
nonzero eigenvalues which coincide with the $\operatorname{rank}(Z)\leq\min
(M,n)$ nonzero eigenvalues of the $n\times n$-matrix $Z^TZ$. This is
equivalent in fMRI to saying that there is a relation between the
covariance matrix of space (a huge $M\times M$ matrix) and that of the
time dimension ($n\times n$ matrix). Furthermore, the eigenvectors
$v_k$ of $Z Z^T$ can be obtained from the eigenvectors $v^{\prime}_k$
of $Z^TZ$ by
\[
v_k=\frac{Z v^{\prime}_k}{\|Z v^{\prime}_k\|},\qquad
k=1,\ldots,\operatorname{rank}(Z).
\]
For more details we refer to \citet{haerdlesimar}, Chapter 8.4.
This indicates that temporal eigenvectors and spatial eigenvectors are
intrinsically linked due to the way the data is discretely collected
with no physical meaning whatsoever. Without presmoothing of the
observed data, it can easily happen that $M\gg n$ (as is the case of
fMRI where the number of voxels usually far exceeds the number of time
points). This implies that a maximum of $n$ different components can be
found. Consequently, even though there are hundreds of thousands of
voxels recorded, only a few hundred components are actually
identifiable, if the analysis proceeds in this generic way. In the case
where $M\gg n$, it is computationally much faster to calculate the
eigenvectors of $Z^TZ$ and then use the above transformation to obtain
the eigenvectors of $ZZ^T$. This computational idea has been used for
magnetic resonance imaging data (anatomical imaging rather than fMRI)
in an i.i.d. setting in \citet{Zipunnikovetal2010}.

\subsection{Separable covariance structures}\label{subsectionsepcov}
The above discussion suggests that in many settings a loss of precision
is unavoidable when the nonparametric covariance estimator (\ref
{eqcovnonparam}) is used with such high-dimensional data. Therefore, in
this section we assume a separable data structure which reduces the
number of unknown parameters and can significantly improve
computational speed as well as accuracy, at least in situations where
the data structure is correctly specified. The use of separable
functions for brain imaging is well known, either for smoothing
[\citet{Worsleyetal2002}] or signal processing using techniques
such as separable wavelets [\citet{Ruttimann98}], both of which
indirectly imply separable covariances.

As well as having been previously suggested for multivariate
covariances for images [see \citet{DrydenBBS2009} for an example and
related references], separable covariance structures have obtained
significant attention in the context of spatio-temporal statistics,
where they have been used to separate the purely temporal covariance
from the purely spatial covariance [see \citet{fuentes2004} and \citet
{mitchelletal2005}]. While in our setup a temporal dependency is also
present, we use the separability approach only on the multidimensional
spatial structure mainly for computational reasons to get a better and
more stable approximation of the eigenfunctions in situations where the
temporal sample is only moderately sized and the spatial structure is
very high dimensional.

For clarity of explanation, two-dimensional data sets will be discussed
here, although identical arguments apply for any finite number of\vadjust{\goodbreak}
dimensions. Indeed, the fMRI data set we consider is three dimensional
so that a three-dimensional version of the procedure below is used.

To this end, consider the set $\mathcal{U}_1\times\mathcal{U}_2$, which
is a product of two compact sets. Heuristically, these can be thought
of as the two directions in a planar image.
Let $X_t(u_1,u_2),u_1\in\mathcal{U}_1, u_2\in\mathcal{U}_2$,
$t=1,\ldots
,n$, and under $H_0$,
%
\begin{equation}
\label{eqH0sep} X_t(u_1,u_2)=Y_t(u_1,u_2)+
\mu(u_1,u_2),
\end{equation}
where the mean function $\mu(\cdot,\cdot)$ as well as the functional
stationary time series $\{Y_t(\cdot,\cdot)\dvtx1\leq t\leq n\}$ are elements
of $L^2(\mathcal{U}_1\times\mathcal{U}_2)$, $\E Y_t(u_1,u_2)=0$.

The restricted covariance kernel of $Y_1(\cdot,\cdot)$ is assumed to fulfill
%
\begin{equation}
\label{eqSepCov} c\bigl( (u_1,u_2),(s_1,s_2)
\bigr)=c_1(u_1,s_1) c_2(u_2,s_2),
\end{equation}
where $c_1(u_1,s_1)$ is an element of $L^2(\mathcal{U}_1\times
\mathcal
{U}_1)$ and $c_2(u_2,s_2)$ an element of $L^2(\mathcal{U}_2\times
\mathcal{U}_2)$, with the full covariance function being an element of
$L^2((\mathcal{U}_1\times\mathcal{U}_2)\times(\mathcal{U}_1\times
\mathcal{U}_2))$. An important example of random data having such a
separable structure is the following: assume $Y$ has mean 0 and
covariance kernel $c_Y(u_1,s_1)$ independent of $X$, which has mean 0
and covariance kernel $c_X(u_2,s_2)$, then $Z(u_1,u_2)=Y(u_1)X(u_2)$
has covariance kernel $c_Y(u_1,s_1)c_X(u_2,s_2)$. In this example the
data set itself is separable, from which the separability of the
covariance as well as sample covariance kernel follows.

The factors $c_1$ and $c_2$ can only be obtained up to a multiplicative
constant as
\[
c\bigl( (u_1,u_2),(s_1,s_2)
\bigr)=\bigl(\alpha c_1(u_1,s_1)\bigr)
\biggl(\frac{1}{\alpha} c_2(u_2,s_2) \biggr),\qquad
\alpha\neq0,
\]
but this does not cause a problem for the change-point procedures, as
will be seen below.

As in the nonparametric case, one uses a discretized version of the
covariance matrix for computations, so that this approach significantly
reduces the computational complexity. For instance, if the observations
consist of 100 data points in each direction (as is approximately the
case in one slice of an fMRI image), the covariance ``matrix'' $c$ is a
$10\mbox{,}000\times10\mbox{,}000$ matrix, while $c_1$ and $c_2$ are of
dimension
$100\times100$ each. The covariance matrix of a two-dimensional data
set $Z$ can, for example, be obtained as the covariance matrix of
$\widetilde{Z}=\operatorname{vec}(Z)$, where $\operatorname{vec}$ is the operation that
turns matrices into vectors by ``stacking'' the columns. Under the above
separability assumption, the covariance matrix of $\widetilde{Z}$
corresponds to $c=c_1\otimes c_2$, where $\otimes$ is the Kronecker
product. Obviously the gains from this procedure will be even more in a
3-D fMRI image, where the corresponding full covariance will be of the
order $10^5\times10^5$.

Furthermore, several approaches to estimate $c_1$ and $c_2$ from the
data in a multivariate setting have been discussed in the literature.\vadjust{\goodbreak}
\citet{vanloanpitsianis} propose an algorithm which approximates a
possibly nonseparable covariance matrix by the closest (in the
Frobenius norm) Kronecker product which has been shown to be useful in
spatio-temporal covariance matrix approximation [\citet{Genton2007}].
While this is a very appealing approach, especially in view of
misspecification, it is computationally not feasible in a
high-dimensional context, as it involves the calculation of singular
vectors, which is computationally very expensive. \citet{dutilleul99}
proposes an MLE algorithm to estimate the factors, but again for
high-dimensional data it is computationally too slow. However, their
approach is related in the sense that they propose to start their
algorithm with our estimator below. This amounts to our estimator being
asymptotically unbiased but not efficient (although computationally
feasible, which is one of our main requirements). Extended and related
algorithms have also been proposed for the estimation of separable
covariance functions in a signal processing context [\citet
{Werneretal2008}] but are again designed for the use in
small-dimensional problems.

The covariance kernels
%
\begin{equation}
\label{eqestsepcovkern1} c_1(u_1,s_1) =
\int_{\mathcal{U}_2}c\bigl((u_1,z),(s_1,z)
\bigr)\,dz 
\end{equation}
and, equivalently, $c_2(u_2,s_2)$ also need to be estimated from the
discretely observed data. Here we adopt an approach based on the
empirical covariance,
%
\begin{equation}\label{eqsepcovnonparam}
\widehat{c}_1(u_1,s_1)=\int
_{\mathcal{U}_2}\widehat {c}\bigl((u_1,z),(s_1,z)
\bigr)\,dz,
\end{equation}
where $\widehat{c}((u_1,u_2),(s_1,s_2))$ is the multidimensional
analogue of (\ref{eqcovnonparam}). For discretely sampled data (as in
fMRI), the integral is approximated by the following sum:
%
\begin{equation}
\label{eqsepcovnonparamdisc} \frac1 n \sum_{t=1}^n
\frac{1}{|\mathcal{U}_2|}\sum_{z\in\mathcal
{U}_2}\bigl(X_t(u_1,z)-
\bar{X}_n(u_1,z)\bigr) \bigl(X_t(s_1,z)-
\bar{X}_n(s_1,z)\bigr),
\end{equation}
where $\mathcal{U}_2$ in (\ref{eqsepcovnonparamdisc}) is the set of
discrete observations of the function in the second direction (and
where in the 3-D fMRI data, $|\mathcal{U}_1|=64$, $|\mathcal{U}_2|=64$,
and $|\mathcal{U}_3|=33$, yielding a combined $|\mathcal{U}|\approx
135\mbox{,}000$). This approximation amounts to estimating covariances in one
direction while keeping the other directions fixed, and then averaging
over the results. A completely analogous definition for $\widehat
{c}_{2}(u_2,s_2)$ can be used. The individual functions are only
identified up to a multiplicative constant, but the eigenfunctions are
identifiable up to their sign. For details we refer to Section \ref
{secsepproc}, while Table~\ref{tsepcovest} gives an outline of the
overall procedure. This approach not only inherently provides more data
to estimate each set of directional components compared with the
standard approach, but also allows more than the maximum $n$ components
identifiable in the generic nonseparable procedure to be estimated as nonzero.

\begin{table}
\caption{Steps to compute separable eigenfunctions}\label{tsepcovest}
\begin{tabular*}{\tablewidth}{@{\extracolsep{\fill}}lp{0.95\textwidth}@{}}
\hline
1&For each of $k$ dimensions calculate the univariate directional
covariance function with replicates across both time and the other
dimensions. Note that the unidentifiable constants do not matter so can
be set to any arbitrary value,
for example, in two dimensions, the first directional covariance function
is
\[
c_1(u_1,s_1) = \frac1 n \sum_{t=1}^n \frac{1}{|\mathcal
{U}_2|}\sum_{z\in\mathcal{U}_2}\bigl(X_t(u_1,z)-\bar{X}_n(u_1,z)\bigr)\bigl(X_t(s_1,z)-\bar
{X}_n(s_1,z)\bigr).
\]
\\
2&For each directional covariance $i$, $i=1,\ldots,k$, obtain
eigenfunctions $\widehat{v}_{i,j}$ and $\widehat{\lambda}_{i,j}$.\\
3&Order the $\widehat{\lambda}_{i,j}$ and for each $i$, select select
the top $d_i$, for example, $d_i=\sqrt[k]{d}$, eigenfunctions. \\
4&Take the tensor product of the selected eigenfunctions to obtain the
eigenbasis,
\begin{eqnarray*}
&\{\widehat{v}_{1,j_1}\otimes\cdots\otimes\widehat{v}_{1,j_k},
j_l=1,\ldots,d_l, l=1,\ldots, k \}.&\\[-28pt]
\end{eqnarray*}
\\
\hline
\end{tabular*}
\end{table}

In real data, separability can be a somewhat difficult assumption to
verify empirically. However, even if separability is not a valid
assumption, the above procedure still provides a completely valid
projection. The estimated basis functions will just no longer coincide
with the eigenfunctions. However, none of the subsequent methodology
for the change-point model which will be developed in Section~\ref
{secmodel} is limited to principal components, so the procedure
remains useful even in the case of nonseparable data.

\subsection{Separable principal component analysis of the connectome
data}\label{subsectionsepconnect}

\begin{figure}

\includegraphics{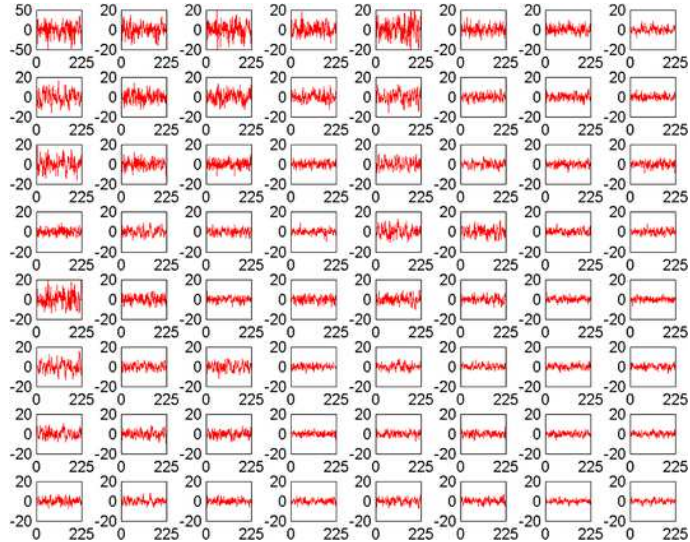}

\caption{Subject 69518: a 64 component functional PCA decomposition of
the brain from this subject (the number of observed spatial locations
is well in excess of 100,000, and thus this represents a massive
dimension reduction). As will be seen later, when testing for
change-points, there was no evidence of an epidemic change for this
subject. This is noticeable in the figure, in that no individual graph
contains sustained deviations in one direction from the mean above
those which might be expected from examining 64 realizations of
stationarity.}\label{figcomponentseriesnull}
\end{figure}

In Figure~\ref{figcomponentseriesnull}, a resting state fMRI data
set is shown after a separable dimension reduction to 64 ($=4\times4
\times4$) dimensions was conducted, using separable projections and
finding the covariance functions using (\ref{eqsepcovnonparam}) from
the previous section. Recall that the original dimensions are
%
$64\times 64\times33$ and therefore more than 2000 times as high. Indeed, the
traditional way of choosing the number of components uses some
threshold for the amount of variance to be explained. For the above
subject (and similarly for the other subjects) 64 components explain
less than 1\% of the variation, which would seem to be of little use in
a dimension reduction context. However, by performing a careful
statistical analysis of the relationship between the type of
change-points to be detected and the choice of the projection, it will
be seen that in many instances, even such a small number of components
will be enough.

\section{Change-point testing and estimation}\label{secmodel}

\subsection{Models for fMRI change-points}\label{sectionCPD}

Activations in brain imaging are typically modeled as changes from
baseline for a short period followed by a return to baseline [see,\vadjust{\goodbreak}
e.g., \citet{Worsleyetal2002}] showing that level shifts or
change-point models describe well the kind of deviation from
stationarity that can be expected. However, in resting state scans, it
is not known when or even if any changes occur across time and, thus,
change-point methods become more applicable than traditional
experimental regression response type models. In addition, epidemic
changes as the simplest model for multiple changes are a good first
approximation to the deviation from stationarity that can be
expected.

The epidemic model is given by
%
\begin{equation}
\label{eqepidemic} X_t(u)=Y_t(u)+\mu(u) +\Delta(u)
1_{\{\vth_1 n<t\leq\vth_2 n\}},
\end{equation}
where $\mu(u)$ is the underlying activation pattern for a particular
subject, and as such does not need to be registered to a standard space
for the model to be evaluated. $Y_t(u)$ is the stationary statistical
deviation from this underlying pattern (it is the stationary covariance
structure of these deviations which are of most interest in
connectivity studies). Here, $\Delta(u)$ is the simplified deviation
from stationarity (a~mean change that persists for a given amount of
the scan, for a fraction $\vth_1$ to $\vth_2$ of the scan, as given by
the $1$ indicator function). Similarly, in a separable situation, the
definition of the model is completely analogous, for example, in two dimensions,
\[
X_t(u_1,u_2)=Y_t(u_1,u_2)+
\mu(u_1,u_2)+\Delta(u_1,u_2)
1_{\{\vth_1 n <
t \leq\vth_2 n\}}.
\]

The epidemic model compares to the AMOC-model, which is given by
%
\begin{equation}
\label{eqAMOC} X_t(u)=Y_t(u)+\mu(u)+\Delta(u)
1_{\{\vth n<t\leq n\}},
\end{equation}
where once the change has occurred it persists for the rest of the
scanning session. We believe that in fMRI studies, epidemic models are
more realistic, but analogous versions of all the results of the paper
are equally valid for AMOC models.

We are interested in testing the null hypothesis of no change in the mean
\[
H_0\dvtx \E X_t(\cdot)=\mu(\cdot),\qquad t=1,\ldots,n,
\]
versus the epidemic change alternative
\begin{eqnarray}
H_1\mbox{:\quad}\E X_t(\cdot)&=&\mu(\cdot),\qquad t=1,\ldots, \lfloor
\vth_1 n\rfloor,\lfloor\vth_2 n\rfloor+1,\ldots, n,\qquad
\mbox{but}
\nonumber\\
\E X_t(\cdot)&=&\mu(\cdot)+\Delta(\cdot)
\neq\mu(\cdot),\nonumber\\
&&\eqntext{t=\lfloor
\vth_1 n\rfloor+1,\ldots, \lfloor\vth_2 n\rfloor,\qquad
0<\vth_1<\vth_2<1.}
\end{eqnarray}
The null hypothesis corresponds to the cases where $\vth_1=\vth_2=1$.

The setting for independent (functional) observations with AMOC was
investigated by \citet{berkesetal09} as well as \citet{aueetal09b} and
for specific weak dependent processes by \citet{hoerkok09}. We will also
allow for dependency (in time) of the functional observations and focus
on the model with an epidemic change, where after a certain time the
mean changes back. For this model some theoretical results relating to
the detection and estimation of changes are given in \citet{AK2}. The
required mathematical setup for the problem is given in Section S.1 of
the supplementary material [\citet{AKsupp}].

\subsection{Projections under the null and alternative hypotheses}

In classical statistical situations, dimension reduction using
principal components is useful because it maximizes the variance
explained by the projection. In the change-point situation, principal
components are also especially suitable but for completely different
reasons. Heuristically speaking, standard variance estimators (such as
the sample variance) increase in the presence of level shifts.
Similarly, the variance estimate for linear combinations of components
in the multivariate situation based on empirical covariances will
increase if a change is present in the linear combination. Thus, under
the alternative, the principal components of the estimated covariance
matrix will likely contain a change [indicating that assumption (\ref
{eqnonorth}), given later, is fulfilled].

\begin{figure}

\includegraphics{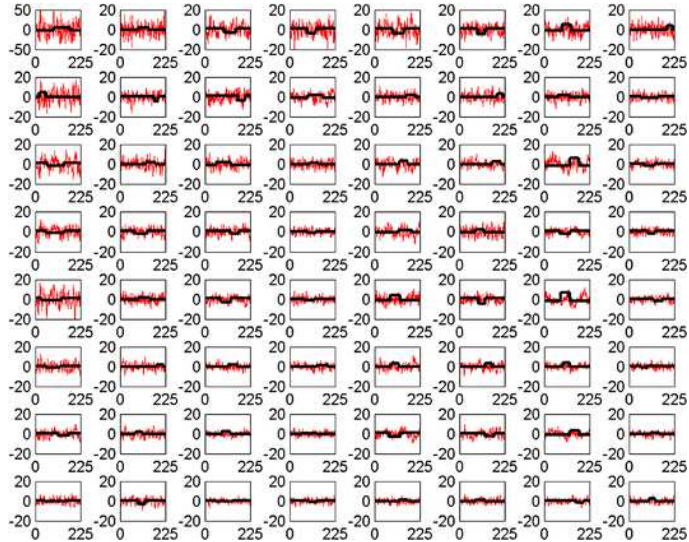}

\caption{Subject 01018: dimension reduction along with possible
epidemic changes indicated (thick black line). Using the tests
described, this subject was found to have deviations from stationarity,
$p<0.001$, even when corrected for multiple comparisons using FDR. This
is most clearly seen in that several of the individual graphs have
large possible sustained deviations in one direction from the
mean.}\label{figcomponentseriesstrong}
\end{figure}

The subject in Figure~\ref{figcomponentseriesstrong} seems to
exhibit strong deviations from stationarity---in fact, the $p$-value
associated with this subject is below $0.001$ based on the bootstrap
test given in Section~\ref{secfMRIboot}.\vadjust{\goodbreak} It should be stressed that
the change detection is a global hypothesis test combined over all
components considered. In this way, while taking more components will
help increase the chance that the change is present in one, it will
come at the cost of the size of the change needed in finite samples for
an omnibus test of this type. However, the subject shown in the figure
did cause a rejection of the null hypothesis of no change both in the
64 and 125 subspace size omnibus tests. While the pictures in Figure
\ref{figcomponentseriesstrong} indicate that an epidemic change is
indeed a good first approximation for the nonstationarities occurring
for this particular subject, more deviation (maybe more change-points)
does seem to be present.
In Figure~\ref{figcomponentseriesweak}, a second subject is shown
with a much smaller deviation from stationarity (most of the components
seem to have little to no possible mean change present), which is
significant but does not survive the false discovery rate (FDR)
correction (see Section~\ref{ssteststat}).

\subsection{Test statistic and estimator of change-point
locations}\label{ssteststat}

For a $d$-dimen\-sional subspace projection, \citet{AK2} propose
to use the following standard change-point statistics\vspace*{1pt} for
an epidemic change on the projected data
$\widehat{\bolds{\eta}}_t=(\widehat{\eta
}_{t,1},\ldots,\widehat{\eta}_{t,d})^T$:
%
\begin{eqnarray}
\label{eqstat}
T_n^{(A)}&=& \frac{1 }{n^3} \sum
_{1\leq k_1<k_2\leq n}\mathbf{S}_n ({k_1}/{n},{k_2}/{n}
)^T\widehat{\bolds{\Sigma }}{}^{-1}\mathbf
{S}_n ({k_1}/{n},{k_2}/{n} ),
\nonumber\\[-8pt]\\[-8pt]
T_n^{(B)}&=&\max_{1\leq k_1<k_2\leq n} \frac{1}{n}
\mathbf{S}_n ({k_1}/{n},{k_2}/{n}
)^T\widehat{\bolds{\Sigma }}{}^{-1}\mathbf
{S}_n ({k_1}/{n},{k_2}/{n} ),
\nonumber
\end{eqnarray}
where $\widehat{\bolds{\Sigma}}$ is a consistent estimator for the
long-run covariance matrix [as defined in (\ref{eqdeflrcov})] and
\[
\mathbf{S}_{n}(x,y)= \sum_{nx< j \leq ny} \Biggl(
\widehat{\bolds{\eta }}_j-\frac1 n\sum
_{t=1}^n\widehat{\bolds{\eta}}_t
\Biggr).
\]
For the small sample performance of the test the choice of estimator
$\widehat{\bolds{\Sigma}}$ is crucial, which is why this issue is
discussed in detail in Section~\ref{remflattop}.

\begin{figure}

\includegraphics{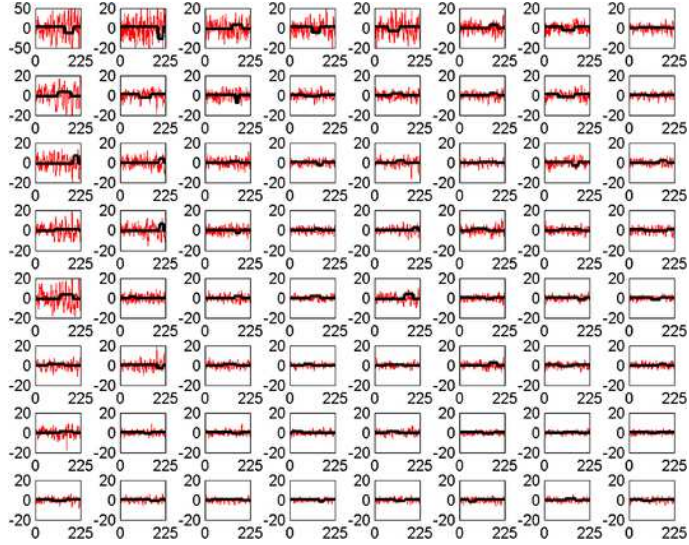}

\caption{Subject 48501: dimension reduction along with possible
epidemic changes indicated (thick black line). Using the tests
described, this subject was found to have deviations from stationarity,
$p<0.05$, but not evidence of deviations when using FDR multiple
comparisons correction. In this case some individual graphs seem to
show more evidence of mean change than in Figure
\protect\ref{figcomponentseriesnull} but less coherence in terms of time
than some in Figure
\protect\ref{figcomponentseriesstrong}.}\label{figcomponentseriesweak}
\end{figure}

The main aim of the test statistics above is to determine regions where
the mean differs significantly from the overall mean of the complete
time series. If these differences are larger than a threshold, then a
change-point is deemed to have occurred. The limit distributions of the
statistics will be found in Section~\ref{secasymptotics}. If the value
is above the threshold, then in the same way as with many CUSUM type
change-point tests, good estimators are usually obtained for the
change-point locations by taking the points where the statistics
achieve their maximum. Thus, as an estimator for the change-points, we propose
%
\begin{equation}
\label{eqestepidemic} (\widehat{\vth}_1,\widehat{
\vth}_2)=\arg\max \bigl(\mathbf {S}_n^T(x,y)
\widehat{\bolds{\Sigma}}{}^{-1}_n\mathbf
{S}_n(x,y)\dvtx0\leq x<y\leq1 \bigr),
\end{equation}
where $\mathbf{S}_n(x,y)$ is as above and $(x_1,y_1)=\arg\max
(Z(x,y)\dvtx0\leq x<y\leq1)$ iff $x_1=\min(0\leq x<1\dvtx Z(x,y)=\max_{0\leq
s<t\leq1}Z(s,t)$ for some $y)$ and $y_1=\max(y>x_1\dvtx Z(x_1,y)=\max_{0\leq
s<t\leq1}Z(s,t))$.

\begin{figure}
\begin{tabular}{@{}c@{}}

\includegraphics{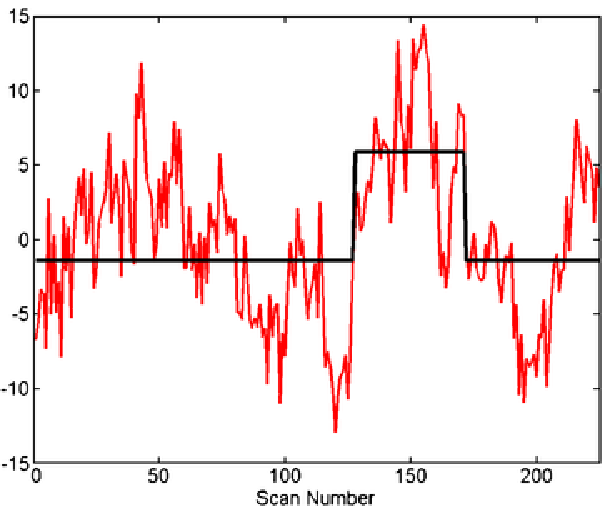}
\\
\textup{(a)}\\[6pt]

\includegraphics{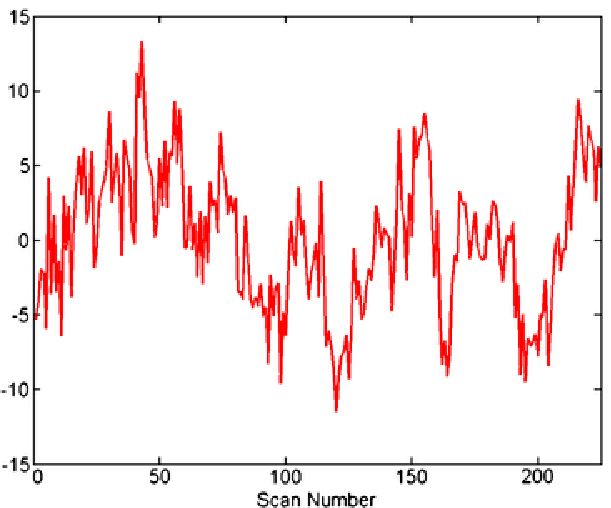}
\\
\textup{(b)}
\end{tabular}
\caption{Subject 01018: for this subject there is evidence of
deviations from stationarity, $p<0.001$.
This figure shows a candidate component 23 time series \textup{(a)}
before and \textup{(b)}
after correction using the estimated change-point location.}\label
{figcomponentseriesstrongsingle}
\end{figure}

\begin{figure}
\begin{tabular}{@{}c@{}}

\includegraphics{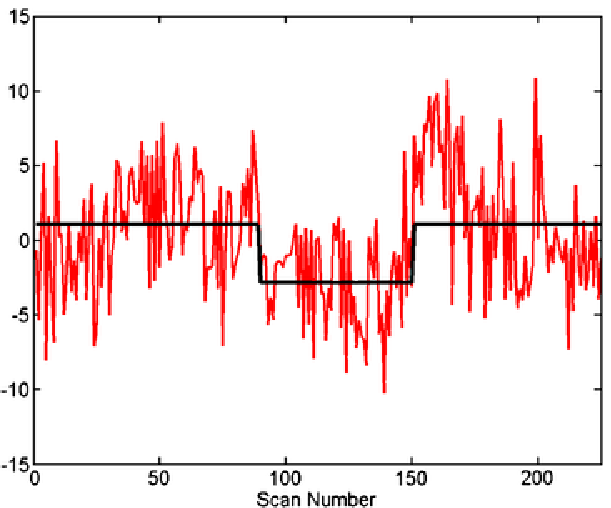}
\\
\textup{(a)}\\[6pt]

\includegraphics{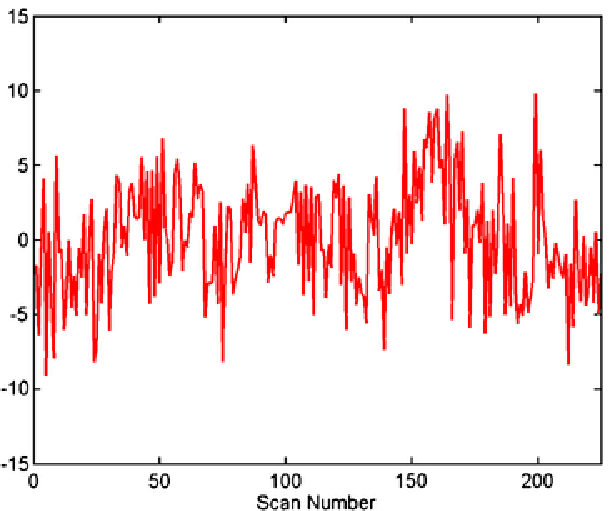}
\\
\textup{(b)}
\end{tabular}
\caption{Subject 48501: for this subject there is evidence of
deviations from stationarity, $p<0.05$,
but it is no longer rejected when using FDR multiple comparisons correction.
This figure shows a candidate component 7 time series \textup{(a)}
before and
\textup{(b)} after correction using the estimated change-point location.}
\label{figcomponentseriesweaksingle}
\end{figure}

\begin{figure}
\begin{tabular}{@{}c@{}}

\includegraphics{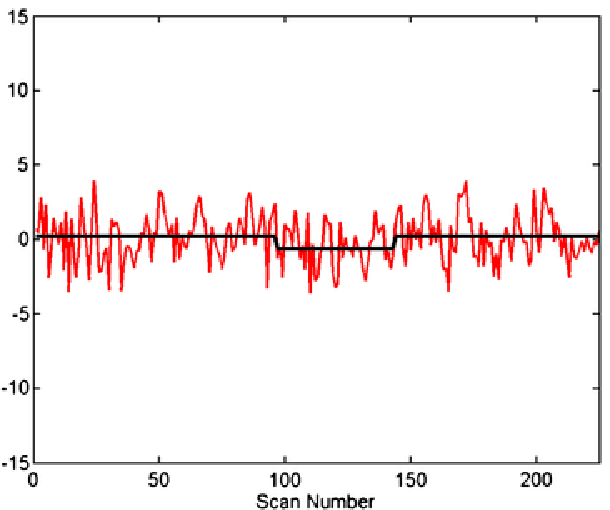}
\\
\textup{(a)}\\[3pt]

\includegraphics{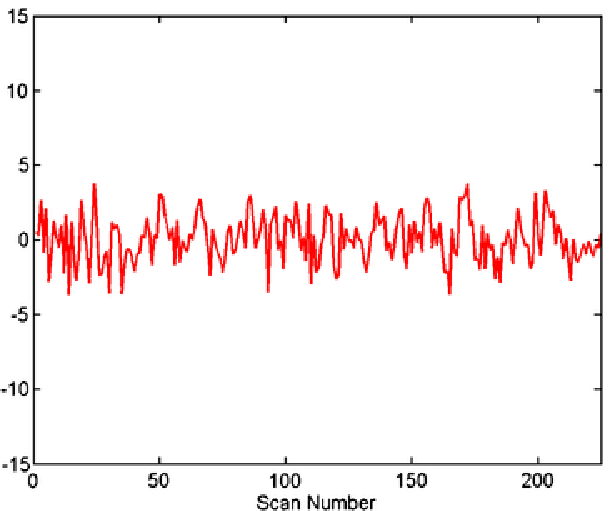}
\\
\textup{(b)}
\end{tabular}
\caption{Subject 48501: as mentioned above, for this subject there is
evidence of weak deviations from stationarity, $p<0.05$, but not
rejected when using FDR multiple comparisons correction. This figure
shows one of the components from the subject that has little evidence
of any kind of nonstationarity present. While the black line results
from the estimator from the maximum of the change-point statistic, it
is not a viable candidate series to contain nonstationarities.
\textup{(a)} Component 56 time series; \textup{(b)} component 56:
epidemic change removed.}\label{figcomponentseriesweakstationary}
\end{figure}

Figures~\ref{figcomponentseriesstrongsingle}--\ref
{figcomponentseriesweakstationary} show three component time series
selected for their different properties. The component in Figure \ref
{figcomponentseriesstrongsingle} can be seen to be a candidate
series for a change to have occurred with the resulting change
corrected series visually appearing much more stationary (although it
is likely there are other nonstationarities present as well). This
series, from subject 01018 in the connectome data set, was found to
have evidence of nonstationarities when the sample version of the
statistic (given in Section~\ref{sectionbootstrap}) was tested on both
a 64 and 125 component projection.

When testing subject 48501 from the connectome data, from whom the
components can be seen in Figure~\ref{figcomponentseriesweak}, an
epidemic change seems to be quite a good model for several components,
but only a small part of the time series deviates from stationarity.
For example, component 7 in Figure \ref
{figcomponentseriesweaksingle} shows a less pronounced but still
plausible epidemic change compared with component 23 of subject 01018
in Figure~\ref{figcomponentseriesstrongsingle}. However, as can be
seen in another component (Figure \ref
{figcomponentseriesweakstationary}) from subject~48501, some of the
components seem to be stationary without any change present.

Given that nearly 200 subjects were tested, a multiple comparison
correction was implemented using the independent FDR method by \citet
{BenjaminiHochberg}. The use of an independent FDR is based on the fact
that the comparisons are being taken across subjects who can be assumed
to be independent of each other. Subject 01018 (Figure \ref
{figcomponentseriesstrong}) survived the FDR correction and evidence
was still found of nonstationarities being present. Subject 48501,
whose projections are seen in Figure~\ref{figcomponentseriesweak},
also rejected the null hypothesis but only at about a 3\% level, hence
not surviving the FDR correction.

Finally, in Figure~\ref{figcomponentseriesnull} the subject shown
has components which do not indicate level shifts and, in fact, the
null hypothesis is not rejected for this subject, either with or
without FDR correction.

\subsection{Questions associated with the application of the above
procedures to fMRI data}

While the discussion above provides a procedure for obtaining test
statistics for functional\vadjust{\goodbreak} data in very high-dimensional settings such
as fMRI, it naturally leads to a number of questions, which the
remainder of the paper seeks to address. These questions and the
sections where they are addressed are as follows:
\begin{longlist}[(5)]
\item[(1)] Could the projected data exhibit a different type of alternative
than the one we are looking for in the full functional time series. In
particular, if there is an epidemic change in the fMRI data, will there
still be an epidemic change in any nonstationary component derived from
the projection? (See Section~\ref{secaltprojs}.)
\item[(2)] Is there a limit distribution available for these test statistics
under the null hypothesis such that critical values can be obtained so
that deviations from stationarity can be determined for fMRI? What
happens to the statistics under the alternative hypothesis, that is,
when there are nonstationary portions in the brain activity? (See
Section~\ref{secasymptotics}.)
\item[(3)] How is the power of the test (possibility of detecting changes)
related to the projection that is taken? This is critical if only a
small number of components can feasibly be taken, as in the case of
fMRI, where computational considerations will dominate. (See Section
\ref{secpowsep}.)
\item[(4)] Can this all be done when there are only relatively small samples
of functional data available (an fMRI time series is typically only a
few hundred time points with hundreds of thousands of spatial
locations)? (See Section~\ref{sectionbootstrap}.)
\item[(5)] As most fMRI studies have multiple subjects, can information
about change-points be generalized to the population? (See Section \ref
{sectionhierarchical}.)\
\end{longlist}

\section{Some statistical properties of the
test statistics}\label{secstatsprops}

\subsection{Projections under stationarity and level shifts}
\label{secaltprojs}

The entire brain covariance structure in an fMRI data set, as
represented by the covariance kernel $c(u,s)$, is not known and needs
to be estimated. However, even if $c(u,s)$ were known, using estimators
would often be preferable due to the nice property that the estimated
covariance can be influenced by the change in such a way that the
change becomes detectable in a lower-dimensional projection (cf.
Corollary~\ref{lemseppower}). Thus, even if we knew how the brain
varied in a stationary condition, it would be preferable to use
estimators unless we knew exactly in which lower-dimensional
projections the changes will have occurred (or if, as in a region based
analysis, we are only interested in the stationarity of the projection
rather than the full data). Thus, we need to examine the behavior of
projections under the alternative.

Under an epidemic change alternative ($ t=1,\ldots,n, l=1,\ldots,d$),
%
\begin{equation}
\label{eqscores}\quad \widehat{\eta}_{t,l}:=\langle X_t,
\widehat{v}_l\rangle=\int X_t(u)\widehat{v}_l(u)
\,du=\langle Y_t,\widehat v_l\rangle+1_{\{\vth_1
n < t \leq\vth_2 n\}}
\langle\bolds{\Delta},\widehat v_l\rangle.
\end{equation}
In particular, $\widehat{\bolds{\eta}}_{t}=(\widehat{\eta
}_{t,1},\ldots,\widehat{\eta}_{t,d})^T$ is a $d$-dimensional time
series exhibiting the same type of level shifts, that is, an epidemic
change in this case, as the functional sequence $\{X_t(\cdot)\dvtx1\leq
t\leq
n\}$ if the change is not orthogonal to the subspace spanned by
$\widehat{v}_1(\cdot),\ldots,\widehat{v}_d(\cdot)$.

From (\ref{eqscores}) it is obvious that the choice of estimation
procedure for basis functions has a substantial influence under the
alternative on the size of $\langle\bolds{\Delta},\widehat
v_l\rangle$, hence the visibility and detectability of the change. In
other words, the behavior of this estimation procedure under
alternatives is crucial for the power of the test. As a contrast, the
estimation procedure has only a very mild influence on the behavior
under the null hypothesis.

Under the null hypothesis, we require the estimated orthonormal system
(ON-system) $\{\widehat{v}_l(\cdot),l=1,\ldots,d\}$ (assuming $d$
distinct eigenvalues) to stabilize in the following sense for technical reasons:
%
\begin{equation}
\label{eqON1} \int\bigl(\widehat{v}_l(u)-s_l
v_l(u)\bigr)^2 \,du=O_P\bigl(n^{-1}
\bigr),
\end{equation}
where $s_l=\operatorname{sgn} ( \int v_{l}(u)\widehat{v}_{l}(u) \,du
)$ and $\{v_l(\cdot),l=1,\ldots,d\}$ is some orthonormal system.
In particular, $\{\widehat{v}_l(\cdot),l=1,\ldots,d\}$ is a consistent
estimator of $\{v_l(\cdot),l=1,\ldots,d\}$ up to the sign. In addition,
if the basis is fixed, as in a wavelet based or region based analysis,
this proposition is fulfilled by definition.

It cannot, in general, be expected that the same limit of the estimated
eigenfunctions will occur under both the null and alternative
hypothesis. However, having different limits can actually be favorable
when detecting changes, as will be seen in Corollary \ref
{lemseppower}. Thus, under the alternative we require that
%
\begin{equation}
\label{eqON2} \int\bigl(\widehat{v}_l(u)-s_l
w_l(u)\bigr)^2 \,du=o_P(1),
\end{equation}
where $\{w_{l}(\cdot),l=1,\ldots,d\}$ is an orthonormal system, $\{
\widehat{v}_{l}(\cdot),l=1,\ldots,d\}$ the same estimators as before
and $s_l=\operatorname{sgn} ( \int w_{l}(u)\widehat{v}_{l}(u) \,du  )$,
that is, the estimators converge to some contaminated ON-system. Note
that $w_l$ usually depends on the alternative. Indeed, most statistical
procedures, including PCA, will still have stable behavior even in the
presence of nonstationarities.

None of the above properties require the basis to be the principal
component basis. However, as will be seen in Section \ref
{secPCAstats}, PCA does indeed fulfil the properties (\ref{eqON1})
and (\ref{eqON2}) given above.

\subsection{Asymptotic evaluation}\label{secasymptotics}

Under (\ref{eqON1}) and the time series assumptions given in Section
S.1.2, and where in (\ref{eqstat}) the long run
covariance is defined to be
%
\begin{equation}
\label{eqdeflrcov} \bolds{\Sigma}=\sum_{k\in\mathbb{Z}}
\Gamma(k), \qquad\Gamma (h)=\E \bolds{\eta}_t\bolds{
\eta}_{t+h}^T
\end{equation}
for $h\geq0$, and $\Gamma(h)=\Gamma(-h)^T$ for $h<0$, \citet{AK2} prove
the following asymptotics under $H_0$:
%
\begin{eqnarray}
\label{eqlimitstat}
T_n^{(A)} &\dto&\sum
_{1\leq l\leq d}\iint_{0\leq x<y\leq
1}\bigl(B_l(x)-B_l(y)
\bigr)^2 \,dx \,dy,
\nonumber\\[-8pt]\\[-8pt]
T_n^{(B)}&\dto&\sup_{0\leq x<y \leq1}\sum
_{1\leq l\leq
d}\bigl(B_l(x)-B_l(y)
\bigr)^2,
\nonumber
\end{eqnarray}
where $B_l(\cdot)$, $l=1,\ldots,d$, are independent standard Brownian bridges.

In order to obtain asymptotic power one for the above tests, the
estimation procedure additionally needs to stabilize under
alternatives, as in (\ref{eqON2}). The change can only be detected if
it is not orthogonal to the contaminated ON-system, that is, for some
$k=1,\ldots,d$ it holds
%
\begin{equation}
\label{eqnonorth} \int\Delta(u)w_{k}(u) \,du\neq0.
\end{equation}
Then, \citet{AK2} show that under the epidemic change alternative
\[
T_n^{(A)}\pto\infty,\qquad T_n^{(B)}\pto
\infty,
\]
if $\widehat{\bolds{\Sigma}}\pto\bolds{\Sigma}_A$ for some
symmetric positive-definite matrix $\bolds{\Sigma}_A$. This shows
that the power of the test is mostly affected by the estimation
procedure to obtain the orthonormal basis for the projection.

\citet{AK2} prove that the change-point estimator related to the above
test as given in (\ref{eqestepidemic}) is consistent under the
assumptions in Section~S.1.1 and even get the
following rate given slightly stronger assumptions:
%
\begin{equation}
\label{eqestrate} (\widehat{\vth}_1,\widehat{\vth}_2)-(
\vth_1,\vth_2)=O_P\bigl(n^{-1/2}
\bigr).
\end{equation}

\subsection{Specifics for principal component analysis}\label{secPCAstats}

When using PCA, the basis is defined from the data via the empirical
covariance function. Thus, the properties of the empirical estimator of
the covariance are important. In order to get (\ref{eqON1}), we
require that the estimated covariance kernel $\widehat{c}_n(u,s)$ is a
consistent estimator for the covariance kernel $c(u,s)$ of $\{Y_1(\cdot
)\}$ with convergence rate $\sqrt{n}$ under $H_0$, that is,
%
\begin{equation}
\label{assCnull} \iint\bigl(\widehat{c}_n(u,s)-c(u,s)
\bigr)^2 \,du \,ds=O_P\bigl(n^{-1}\bigr).
\end{equation}

\citet{AK2} show that strong mixing and other weak dependent sequences
fulfill this assumption. This condition implies that (\ref{eqON1})
holds for standard PCA, with $v_l(u)$ being the associated principal
components [\citet{AK2}]. The equivalent result for separable PCA will
be discussed in the next section.

Under the alternative $H_1$, we assume that there exists a covariance
kernel $k(u,s)$, such that
%
\begin{equation}
\label{assCalt} \iint\bigl(\widehat{c}_n(u,s)-k(u,s)
\bigr)^2 \,du \,ds\pto0,
\end{equation}
which similarly implies that (\ref{eqON2}) holds, with $w_l(u)$ being
the associated principal components [\citet{AK2}].

In case of independent functional observations and for an AMOC change
alternative, \citet{berkesetal09} proved (\ref{assCnull}) as well as
(\ref{assCalt}) for the estimator for the covariance given in (\ref
{eqcontkernel}). Their proof can be extended to the dependent AMOC
situation [cf. \citet{hoerkok09}] as well as the dependent epidemic
change situation [cf. \citet{AK2}]. For the latter the contaminated
covariance kernel is given by
%
\begin{equation}
\label{eqcontkernel} k(u,s)=c(u,s)+\theta(1-\theta)\Delta(u) \Delta(s),
\qquad \theta =
\vth_2-\vth_1>0.
\end{equation}
In particular, this shows that there will be a systematic error if the
covariance structure is estimated with level shifts present. For fMRI
resting state studies where estimating connectivity is the major aim,
this amounts to detecting false correlations which are not related to
the true connectivity, as measures of connectivity will be derived from
$k$ in any subsequent correlation analysis rather than the true $c$ covariance.

The above discussion shows that the contaminated covariance kernel
$k(u,s)$ as well as the contaminated eigenvalues $\gamma_l$ will
usually depend on the type and shape of the change. Interestingly, for
$k$ as in (\ref{eqcontkernel}), this is a feature rather than a
problem, which leads to the desirable property that a large enough
change can influence $k$ in such a way that it automatically is not
orthogonal to the chosen subspace if the eigenfunctions belonging to
the largest eigenvalues of $\widehat{c}_n$ are used (cf.
Corollary~\ref{lemseppower} as well as
Theorem S.2.2 in the supplementary material
[\citet{AKsupp}]).

\subsubsection{Separable projections}\label{secsepproc}

If the covariance kernel is indeed separable, use of a separable
estimator leads to a correct estimation of the noncontaminated
eigenspace under $H_0$ and to the estimation of a well-defined
contaminated eigenspace under $H_1$. However, even in the misspecified
case, that is, when the covariance kernel has no separable structure,
one estimates the basis functions of a well-defined subspace under both
$H_0$ as well as $H_1$ but with a different interpretation (cf.
Theorem S.2.1 in the electronic supplementary material
[\citet{AKsupp}]).

The eigenvalues $\lambda_l$, respectively, functions $v_l$ corresponding to a
separable $c$ are the products of the eigenvalues $\lambda_{1,i},
\lambda_{2,j}$, respectively, functions $v_{1,i}, v_{2,j}$ of $c_1$ and $c_2$,
since by (\ref{eqdefeigen})
%
\begin{eqnarray}
\label{eqsepeigenv}
&&
\int_{\mathcal{U}_1}\int_{\mathcal{U}_2} c
\bigl( (u_1,u_2),(s_1,s_2)
\bigr) v_{1,i}(s_1)v_{2,j}(s_2)
\,ds_1 \,ds_2
\nonumber
\\
&&\qquad= \int_{\mathcal{U}_1}c_1(u_1,s_1)v_{1,i}(s_1)
\,ds_1 \int_{\mathcal
{U}_2} c_2(u_2,s_2)v_{2,j}(s_2)
\,ds_2
\\
&&\qquad=\lambda_{1,i} \lambda_{2,j} v_{1,i}(u_1)
v_{2,j}(u_2).
\nonumber
\end{eqnarray}
We propose to use the subspace spanned by the first $d_1$ principal
components of $c_1$ in the first dimension and the first $d_2$
principal components of $c_2$ in the second dimension. In a balanced
situation it makes sense to choose $d_1=d_2$, but sometimes there are
fewer observations in one direction after discretization in which case
$d_1\neq d_2$ may be preferable. This balanced choice of basis
selection is preferable to choosing a basis of the eigenfunctions
belonging to the largest $d$ joint eigenvalues, as only then the
eigenfunction will be guaranteed to include a large enough separable
change (cf. Remark S.2.1 in the electronic
supplementary material
[\citet{AKsupp}]). 

The empirical covariance kernel $\widehat{c}_n (
(u_1,u_2),(s_1,s_2)  )$ as in (\ref{eqcovnonparam}) is used to
estimate $c_1$ and $c_2$ as in (\ref{eqsepcovnonparam}). In case of
separability of $c$ it holds
\[
\widehat{c}_j(u_j,s_j) \pto
\frac{\tr c}{\tr c_j} {c}_j(u_j,s_j),\qquad j=1,2,
\]
where $\tr c(x,y)=\int c(x,x) \,dx$ and $\tr c=\sum_{i\geq1}\lambda_i>0$, if $c\neq0$, where $\lambda_i$ are the eigenvalues of the
covariance operator $C v=\int_{\mathcal{U}_1\times\mathcal{U}_2}
c(\cdot
,y)v(y) \,dy$ [cf. Theorem~4.1 in \citet{gohberg03}] and analogously
$\tr c_j>0$.
For the purpose of estimating the $d$ largest principal components,
this additional constant does not make a difference since the
eigenfunctions are the same and the eigenvalues are only multiplied by
a positive constant, thus not changing the order.

Correspondingly, define
%
\begin{equation}
\label{eqsubspacesep}\qquad \widehat{v}_{(r,l)}(u_1,u_2)=
\widehat{v}_{1,r}(u_1)\widehat {v}_{2,l}(u_2),\qquad
r=1,\ldots,d_1,l=1,\ldots,d_2,
\end{equation}
where $ \widehat{v}_{i,r}$ is the $r$th principal component of
$\widehat
{c}_i$ as in (\ref{eqestsepcovkern1}).

To understand the behavior of this estimator under $H_0$ for a
possibly nonseparable $c$, let
%
\begin{eqnarray}
\label{eqkerntilde}
\widetilde{c}_1(u_1,s_1)&=&
\int_{\mathcal
{U}_2}{c}\bigl((u_1,z),(s_1,z)
\bigr)\,dz,\nonumber
\\
\widetilde{c}_2(u_2,s_2) &=& \int
_{\mathcal
{U}_1}{c}\bigl((z,u_2),(z,s_2)\bigr)\,dz,
\\
\widetilde{c} \bigl( (u_1,u_2),(s_1,s_2)
\bigr)&=&\widetilde {c}_1(u_1,s_1)
\widetilde{c}_2(u_2,s_2).
\nonumber
\end{eqnarray}
If the covariance kernel $c$ is separable, that is, fulfills (\ref
{eqSepCov}), then $\widetilde{c}_j=\frac{\tr c}{\tr
c_j}
c_j$, $j=1,2$ and $\widetilde{c}=\tr c c$, that is, the space
spanned by $\widehat{v}_{(r,l)}(u,s)$, $r=1,\ldots,d_1,\allowbreak
l=1,\ldots,d_2$,
is indeed the space spanned by the eigenfunctions of the covariance kernel.

It has been discussed in Section~\ref{secPCAstats} that
(\ref{assCnull}) holds for a wide range of processes, where the
covariance kernel $c$ need not be separable. If the eigenvalues of
$\widetilde{c}$ are identifiable in the sense that
$\widetilde{\lambda}_{i,1}>\widetilde{\lambda}_{i,2}>\cdots
>\widetilde{\lambda}_{i,d_i+1}\geq\widetilde{\lambda
}_{i,d_i+2}\geq
\cdots\,$, $i=1,2$, then, $\widehat{v}_{(r,l)}(u_1,u_2)$ and
$v_{(r,l)}(u_1,u_2)=\widetilde{v}_{1,r}(u_1) \widetilde{v}_{2,l}(u_2)$,
$r=1,\ldots,d_1,l=1,\ldots,d_2$, fulfill (\ref{eqON1}), where
$\widetilde{v}_{i,r}$ is the $r$th principal component of
$\widetilde{c}_i$ (for details we refer to Theorem S.2.1 in
the electronic supplementary material [\citet{AKsupp}]). In particular,
if $c$ is separable, this proves the corresponding consistency result.

Assume that (\ref{assCalt}) holds with a contaminated covariance
kernel $k ( (u_1,u_2)$, $(s_1,s_2)  )$ under the alternative, as
is the case with many weak dependent processes (as discussed in Section
\ref{secPCAstats}). Define $\widetilde{k}_{1}$, $\widetilde{k}_{2}$,
$\widetilde{k}$ based on the contaminated covariance kernel $k(
(u_1,u_2),(s_1,s_2))$ analogously to $\widetilde{c}_{1}$,
$\widetilde{c}_{2}$, $\widetilde{c}$ above. Then, an analogous
assertion to the one of the preceding paragraph holds if one replaces
all covariance kernels correspondingly (for details we refer to
Theorem S.2.1 in the electronic supplementary material
[\citet{AKsupp}]). As a result, a subspace of the eigenspace
$\widetilde{w}_l$ of $\widetilde{k}$ is used for the change-point
procedure (with $\widetilde{w}_{i,l}$ being the associated
eigenfunctions of $\widetilde{k}_i$). Thus, all changes that are not
orthogonal to this (contaminated) subspace are detectable [cf.
(\ref{eqnonorth}) and following lines].\vspace*{1pt}

Intuitively, $\widetilde{c}_{1}$, $\widetilde{c}_{2}$, $\widetilde{c}$
and analogously $\widetilde{k}_{1}$, $\widetilde{k}_{2}$, $\widetilde
{k}$ can be thought of as separable approximations to the covariance
obtained by first integrating along all directions except the one of
interest and then taking the product of these integrated covariances to
obtain the full covariance (this has similarities to obtaining a joint
distribution by taking the product of the marginals). In the case of a
true separable covariance, the approximation is exact, but even in the
case of a truly nonseparable covariance, the resulting eigenbasis from
the separable approximation is still a completely valid basis to
perform change-point detection.

\subsubsection{Power using separable principal component
analysis}\label{secpowsep}

In Section~\ref{secasymptotics} we have seen that changes are detected
if
\[
\int_{\mathcal{U}_1}\int_{\mathcal{U}_2}\Delta(u_1,u_2)\widetilde
{w}_{1,r}(u_1)\widetilde{w}_{2,l}(u_2)\,du_1\,du_2\neq0
\]
for some $1\leq r\leq d_1, 1 \leq l\leq d_2$. If the eigenfunctions are
estimated using (\ref{eqsubspacesep}), then most changes detectable
by $\widetilde{v}{(r,l)}$ will also be detectable by the contaminated
system $\widetilde{w}{(r,l)}$. In addition, most large enough changes
become detectable using the separable estimation procedure from Section~\ref{secsepproc}.
For details we refer to
Theorem S.2.2 in the electronic supplementary
material [\citet{AKsupp}]. Corollary~\ref{lemseppower} shows one
important example of changes having this nice property, namely,
separable changes, for which
$\Delta(u_1,u_2)=\Delta_1(u_1)\Delta_2(u_2)$.
%
\begin{cor}\label{lemseppower} Assume that the change is separable,
that is, $\Delta(u_1,u_2)=\Delta_1(u_1)\Delta_2(u_2)$. In addition,
assume $\int_{\mathcal{U}_1}\int_{\mathcal{U}_2}\Delta^2(u_1,u_2)
\,du_1 \,du_2\neq0$.
\begin{longlist}[(a)]
\item[(a)] Let $\widetilde{v}_{j,r}$ be the $r$th principal component of
$\widetilde{c}_j$ and $\widetilde{w}_{j,r}$ be the $r$th principal
component of $\widetilde{k}_j$ and let analogously to (\ref{eqcontkernel})
\[
k \bigl( (u_1,u_2),(s_1,s_2)
\bigr)=c\bigl( (u_1,u_2),(s_1,s_2)
\bigr)+\theta (1-\theta)\Delta(u_1,u_2)
\Delta(s_1,s_2).
\]
Then, any change that is not orthogonal to the noncontaminated subspace
is detectable:
\begin{eqnarray}
&&\int_{\mathcal{U}_1}\int_{\mathcal{U}_2}
\Delta_1(u_1)\Delta_2(u_2)
\widetilde{v}_{1,r}(u_1) \widetilde{v}_{2,l}(u_2)
\,du_1 \,du_2\neq0\nonumber\\
\eqntext{\mbox{for some }1\leq r\leq
d_1, 1 \leq l\leq d_2,}
\\
&&\quad\implies\quad\int_{\mathcal{U}_1}\int_{\mathcal{U}_2}
\Delta_1(u_1)\Delta_2(u_2)
\widetilde{w}_{1,r}(u_1) \widetilde{w}_{2,l}(u_2)
\,du_1 \,du_2\neq0\nonumber\\
\eqntext{\mbox{for some }1\leq r\leq
d_1, 1 \leq l\leq d_2.}
\end{eqnarray}
\item[(b)] Let $\Delta_D(u_1,u_2)=D\Delta(u_1,u_2)$. Let $\widetilde
{w}_{j,1,D}$, be the normalized first principal components of
$\widetilde{k}_{j,D}$ obtained analogously to (\ref{eqkerntilde}) with
\[
k_{D} \bigl( (u_1,u_2),(s_1,s_2)
\bigr)=c\bigl( (u_1,u_2),(s_1,s_2)
\bigr)+\theta (1-\theta)\Delta_D(u_1,u_2)
\Delta_D(s_1,s_2).
\]
Then, there exists $D_0>0$ such that
\[
\int_{\mathcal{U}_1}\int_{\mathcal{U}_2} \Delta_D(u_1,u_2)
\widetilde {w}_{1,1,D}(u_1)\widetilde{w}_{2,1,D}(u_2)
\,du_1 \,du_2\neq0
\]
for all $|D|\geq D_0$. This shows that any large enough change is detectable.
In this case it even holds as $D\to\infty$
\[
\biggl\llVert \pm\widetilde{w}_{j,1,D}(\cdot) - \frac{\Delta_j(\cdot
)}{\|
\Delta_j(\cdot)\|}\biggr
\rrVert \to0.
\]
\end{longlist}
\end{cor}

The corollary does not require that the true underlying covariance
structure is separable for the statement still to be true. In the
simpler situation of a general covariance structure and standard
nonparametric covariance estimators, an analogous assertion has been
proven by \citet{AK2}. Theorem S.2.2 explains the
situation for the separable estimation procedure for a general change.
In this case, only a weaker result can be obtained.

Furthermore, for practical purposes it is advisable to include all
eigenfunctions obtained by combinations of a fixed number of
eigenfunctions in each dimension as in (\ref{eqsubspacesep}) instead
of choosing the ones belonging to the largest $d$ eigenvalues.
Otherwise, the assertion of Corollary~\ref{lemseppower}(b) can no
longer be guaranteed. But this assertion shows that any large enough
separable change has a tendency to switch the eigenfunctions in such a
way that it becomes detectable, which is a very desirable result. For
more details on this, we refer to Remark S.2.1 in
the electronic supplementary material [\citet{AKsupp}].

It is clear that the choice of $d_1$ and $d_2$ plays an important role
in terms of whether a change is detected or not. In PCA frequently the
number of components is chosen in such a way that $80\%$ of the
variability are explained. However, Corollary~\ref{lemseppower}(b)
suggests that a small number of components is often sufficient and may
even increase the power. This has inherent practical applications for
fMRI. If 80\% variation needed to be accounted for, then a very large
number of components (in excess of 50,000) would be needed, yet the
procedure still detects change-points even with very few components.
While this is somewhat unexpected and counter-intuitive, it is
suggested by the results of this section.

\section{Practical aspects of small sample
testing}\label{sectionpractaspects}

\subsection{Estimation of the temporal covariance matrix}\label{remflattop}
In the case where one deals with independent data and an estimation
procedure that---under the null hypothesis---captures the true
eigenfunctions of the covariance matrix, the long-run covariance matrix
(\ref{eqdeflrcov}) is diagonal. In this case, only the variance of
the scores need to be estimated, which can be found using the estimated
eigenvalues.

On the other hand, if the data are dependent such as in fMRI time
series or one uses the separable estimation procedure on a nonseparable
covariance structure (such as if the separability assumption is not
satisfied in applications), estimation of the long-run covariance
matrix $\bolds{\Sigma}$ as in (\ref{eqdeflrcov}) is critical
for the change-point procedure to yield reasonable results. However,
this is a very difficult task, especially if the dimension of the
projection subspace is large and the time series short---both of which
are true for fMRI. Additional estimation errors arise from the fact
that possible change-points should be removed prior to the estimation
of the covariance matrix, otherwise systematic errors arise. While this
works approximately in the fMRI example, there is still the problem
that the epidemic change alternative is only a very crude approximation
to the true deviations from stationarity that can occur.

Most estimators for the long-run covariance matrix are based on
\[
\widehat{\bolds{\Sigma}}=\sum_{|h|\leq b_n}w_q(h/b_n)
\widehat {\Gamma}(h)
\]
for some appropriate weight function $w_q$ and bandwidth $b_n$
where $\widehat{\Gamma}(\cdot)$ is an estimator for the autocovariance
matrix of the (uncontaminated) projected data vector.
\citet{hoerkok09} prove consistency of this estimator for weakly
dependent data.
\citet{politis05} proposed to use different bandwidths for each entry of
the matrix in addition to an automatic bandwidth selection procedure
for the class of flat-top weight functions, where some additional
modifications guarantee the estimate to be symmetric and positive definite.
We follow his approach but adapt the estimator in such a way that it
takes possible change-points into account, thus improving the power of
the test. For details in the univariate situation we refer to \citet
{huskirchconfstud}.

However, in our analysis of the connectome data set the use of such an
estimator (which is already one of the best choices in general) is
rather problematic because the statistic is weighted with the inverse
of this estimated long-run covariance matrix. While the estimator is
eventually positive-definite, for small samples as in our data example
(225 time points) and a high-dimensional covariance matrix ($64\times
64$ after dimension reduction in our example) the estimation errors add
up and result in as many as thirty percent of the (by definition
positive) eigenvalues being estimated as negative. Using appropriate
cutting techniques [confer \citet{politis05}], one can solve this
problem in principle, but the cutting point will essentially determine
whether the null hypothesis is rejected or not so that no reliable
statistical inference is possible [the cut point essentially determines
the value of the smallest eigenvalue, but this becomes the most
influential one when the inverse is taken in the test statistic
(\ref{eqstat})]. Even using a conservative cutoff point, the null
hypothesis of stationarity was rejected for all subjects in our data
example with such tiny $p$-values as to seriously question the validity
of the results. More details on the above difficulties and possible
solutions can be found in Section S.3 of the
electronic supplementary material [\citet{AKsupp}].

Therefore, we decided to use a slightly different change-point
statistic which only corrects for the long-run variance and not
possible dependencies between components. The limit distribution of
this modified test statistic has still the same shape as in (\ref
{eqlimitstat}), but the Brownian bridges are no longer independent but
rather exhibit the long-run correlation structure of the projected
data. Furthermore, the results on the estimators (\ref
{eqestepidemic}) given in (\ref{eqestrate}) remain true. This
estimator leads to stable and reasonable results, but since the
statistic is no longer asymptotically distribution-free, we need to
introduce bootstrap methods in the next section. Bootstrap methods are
usually unappealing in fMRI due to the large data structures which need
to be handled, but in our methodology, the bootstrap will take place on
the projected components, yielding a computationally demanding yet
still feasible approach. However, should there be no dependence between
components, or the temporal dependence be identical for every
component, as, for example, often assumed in methods based on wavelets
[see \citet{Astonetal2005} and \citet{Morrisetal2011}], then the limit
distribution of the test statistic below becomes asymptotically pivotal
and asymptotic critical values can be used (with the form of the long
run variance changing with the particular assumptions on the time
series properties), making the procedure very fast (on the order of a
few minutes) for an fMRI data set.

To elaborate, we use the test statistics below where $\widehat
{\bolds{\Sigma}}$ in $T_n^{(A)}$, respectively, $T_n^{(B)}$ in (\ref
{eqstat}) are replaced by $\widetilde{\bolds{\Sigma}}$:
%
\begin{eqnarray}
\label{eqstattilde}
\widetilde{T}_n^{(A)}&=&
\frac{1 }{n^3} \sum_{1\leq k_1<k_2\leq
n}\mathbf {S}_n
({k_1}/{n},{k_2}/{n} )^T\widetilde{
\bolds {\Sigma }}^{-1}\mathbf{S}_n ({k_1}/{n},{k_2}/{n}),
\nonumber\\[-8pt]\\[-8pt]
\widetilde{T}_n^{(B)}&=&\max_{1\leq k_1<k_2\leq n}
\frac{1}{n} \mathbf {S}_n ({k_1}/{n},{k_2}/{n}
)^T\widetilde{\bolds {\Sigma }}^{-1}
\mathbf{S}_n ({k_1}/{n},{k_2}/{n} ),
\nonumber
\end{eqnarray}
where
%
\begin{equation}
\label{eqtildesigma} \widetilde{\bolds{\Sigma}}(i,j)= ({\widehat{
\gamma}}_i 1_{\{
i=j\}
})_{i,j=1,\ldots,d},
\end{equation}
$\widehat{\gamma}_i$ as in equation (\ref{estflattop}) below,
is an estimator for the diagonal matrix of long-run variances:
\[
V=({\gamma}_i 1_{\{i=j\}})_{i,j=1,\ldots,d},\qquad
\gamma_i=\sum_{l\in
\mathbb{Z}}\E\eta_{1,i}
\eta_{1+l,i}. 
\]

To obtain such an estimator for the long-run variances, let
\[
(\widehat{m}_{1,l},\widehat{m}_{2,l})=\arg
\max_{k_1,k_2} \Biggl(\Biggl\llvert \sum_{t=k_1}^{k_2}
\widehat{\eta}_{t,l}-\frac{k_2-k_1} n \sum
_{t=1}^n\widehat{\eta}_{t,l} \Biggr\rrvert
\Biggr) 
\]
be the estimated change-points that are estimated separately in each
component and let
%
\begin{eqnarray}
\label{estuncondata}\quad
\widehat{e}_l(j)&=&\widehat{
\eta}_{j,l}-\bar{\widehat{\eta }}_{\widehat
{m}_{1,l},\widehat{m}_{2,l}}1_{\{\widehat{m}_{1,l}<j\leq\widehat
{m}_{2,l}\}}-
\bar{\widehat{\eta}}{}^{\circ}_{\widehat
{m}_{1,l},\widehat
{m}_{2,l}}1_{\{j\leq\widehat{m}_{1,l}\ \mathrm{or}\ \widehat
{m}_{2,l}<j\}
},
\nonumber
\\
\quad\bar{\widehat{\eta}}_{\widehat
{m}_{1,l},\widehat{m}_{2,l}}&=&\frac{1}{\widehat{m}_{2,l}-\widehat
{m}_{1,l}}\sum
_{j=\widehat{m}_{1,l}+1}^{\widehat{m}_{2,l}}\widehat {\eta }_{j,l},
\\\quad\qquad
\bar{\widehat{\eta}}{}^{\circ}_{\widehat
{m}_{1,l},\widehat{m}_{2,l}}&=&\frac{1}{n-\widehat{m}_{2,l}+\widehat
{m}_{1,l}}\sum
_{1\leq j\leq\widehat{m}_{1,l}, \widehat{m}_{2,l}<j\leq
n}\widehat{\eta}_{j,l},
\nonumber
\end{eqnarray}
be the estimated uncontaminated data. Then, we obtain an estimator of
the uncontaminated autocovariances in each dimension as
\[
\widehat{\gamma}_l(h)=\frac1 n\sum_{j=1}^{n-h}
\widehat {e}_l(j)\widehat {e}_l(j+h),\qquad h\geq0,\qquad \widehat{
\gamma}(h)=\widehat{\gamma }(-h),\qquad h<0.
\]
Finally, we obtain the estimator for the long-run variance in the $l$th
component by
%
\begin{equation}
\label{estflattop} \widehat{\gamma}_l^2=\max \Biggl(
\widehat{\gamma}_l(0)+2\sum_{k=1}^{B_{l}}
w(k/B_{l})\widehat{\gamma}_l(k),\frac{1}{n(n-1)}\sum
_{j=1}^n \widehat{e}_l(j)^2
\Biggr)
\end{equation}
with the following flat-top kernel
\[
w(x)= %
\cases{1,&\quad $|x|\leq1/2$,
\cr
2\bigl(1-|x|\bigr),&\quad $1/2<|x|<1$,
\cr
0,&\quad $|x|\geq1$,} %
\]
and the bandwidth $B_l=2\widehat{b}_{l}$, where $\widehat{b}_l$ is the
smallest positive integer such that
\[
\bigl\llvert \widehat{\gamma}_l(\widehat{b}_{l}+j)/
\widehat{\gamma }_{l}(0)\bigr\rrvert <1.4 \sqrt{\log_{10}
n/n}\qquad\mbox{for }j=1,\ldots,3.
\]
The rightmost part of (\ref{estflattop}) in the parenthesis is chosen
to ensure positivity and scale invariance of the estimator.
Under appropriate regularity conditions on $\widetilde{\eta
}_{t,l}=\int
Y_t(u)v_l(u) \,du$, this estimator is consistent under the null
hypothesis and converges to $\sum_{j\geq1}\cov(\widetilde{\eta
}_{0,l},\widetilde{\eta}_{j,l})$ under alternatives. For a thorough
proof for the simpler one-dimensional problem see \citet{huskirchconfstud}.

\subsection{Resampling procedures for the testing problem}\label
{sectionbootstrap}
Using resampling methods to obtain critical values often leads to
improvements in the size and power of the tests in small samples. In
case of a nonpivotal limit distribution as, for example, when using the
statistics $\widetilde{T}_n^{(A/B)}$ as in (\ref{eqstattilde}),
asymptotic critical values differ from one time series to another so
that resampling methods are the only way to obtain them. This in effect
means that for fMRI data, the critical values are subject specific, as
we are not assuming that the time series dependencies between scans are
the same for all subjects, but in fact we allow them to vary not only
just in a parameter but structurally as well. For applications of the
bootstrap to univariate change-point tests for dependent data we refer
to \citet{kirchblock} and \citet{kirchpolitis}.

In order to keep the procedure simple, we propose to use the following
studentized circular block bootstrap (to allow for the time series
error structure), taking a possible change-point separately in each
component into account:

Let $K$ be such that $n=KL$, $K,L\to\infty$, $K/L\to0$.
\begin{longlist}[(7)]
\item[(1)] Let $\widehat{e}_{l}(j)$ be as in (\ref{estuncondata}).
%
\item[(2)] Draw $U(1), \ldots,U(L)$ i.i.d., independent of $\{X(\cdot)\}$,
such that $P(U(1)=t)=1/n$, $t=0,\ldots,n-1$.
\item[(3)] Let $e_l^*(Kj+k):=\widehat{e}_l(U(j)+k)$, $l=1,\ldots,d$, where
$\widehat{e}_l(j)=\widehat{e}_l(j-n)$ if $j>n$.
\item[(4)] Calculate
\begin{eqnarray*}
{T}_n^{(1)} :\!&=& \frac{1 }{n^3} \sum
_{1\leq k_1<k_2\leq n}\mathbf {S}^*_n ({k_1}/{n},{k_2}/{n}
)^T\widetilde{\bolds {\Sigma }}^{* -1}
\mathbf{S}^*_n ({k_1}/{n},{k_2}/{n} ),
\\
\mathbf{S}^*_n(x,y) &=& \bigl(S_n^*(1),
\ldots,S_n^*(d)\bigr)^T,\\
S_n^*(l)&=& \sum
_{nx< j \leq ny} \bigl(e^*_l(j)-
\bar{e}_n^*(l) \bigr), \\
\bar {e}_n^*(l)&=&\frac1 n\sum
_{t=1}^ne^*_l(t),
\\
\widetilde{\bolds{\Sigma}}^*(i,i)&=&\frac1 n \sum
_{l=1}^{L-1} \Biggl(\sum_{k=1}^K
\bigl(e_i^* ( Kl+k )- \bar{e}_n^*(i)\bigr)
\Biggr)^2, \\
\widetilde{\bolds{\Sigma}}^*(i,j)&=&0 \qquad\mbox{for }i\neq
j,%
\end{eqnarray*}
in case one wants to use statistic $\widetilde{T}_n^{(A)}$ and
analogous versions for different statistics. Mark that the variance
estimators used for the bootstrap are the block sample variances, hence
give the true variances of the conditional bootstrap distribution.
%
\item[(5)] Repeat steps (2)--(4) $M$ times (e.g., $M=1000$).
\item[(6)]$c^*(\alpha)$ is obtained as the upper $\alpha$-quantile of
${T}_n^{(1)},\ldots,{T}_n^{(M)}$.
\item[(7)] Reject if $T_n>c^*(\alpha)$, where $T_n$ is the statistic of
interest, that is, $\widetilde{T}_n^{(A)}$ in the above example, where
one uses the estimator $\widetilde{\bolds{\Sigma}}$ as given in
(\ref{eqtildesigma}).
\end{longlist}
A similar bootstrap has been applied by \citet{huskirchconf} and \citet
{huskirchconfstud} in the univariate situation to obtain confidence
intervals for the change-point. A proof for the validity of the
univariate bootstrap (not taking possible changes into account) in the
nonstudentized case can be found in \citet{kirchdiss} under appropriate
moment assumptions; extensions to the studentized case are immediate
from (4.4) in \citet{huskirchconfstud}. Extensions to the multivariate
situation can be obtained along the same lines using Wold's theorem. An
additional problem in the situation in this paper is that $\widetilde
{\eta}_{t,l}$ is not observed but needs to be estimated. Since only
moment conditions of $\widetilde{\eta}_{t,l}$ are required for the
proofs, extensions to $\widehat{\eta}_{t,l}$ are straightforward.

The choice of the block-length $K$ is difficult---as a rule of thumb,
we propose to use $n^{1/3}$, because a block length of this order
asymptotically minimizes the mean squared error of the corresponding
bootstrap variance estimate for the sample mean [\citet{Lahiri03}, page
39, Theorem 5.4], which is closely related to our situation.

\section{Testing for epidemic changes
in scans within the connectome data set}\label{secfMRIboot}

As discussed in Section~\ref{remflattop}, obtaining a good estimate of
the full long-run covariance matrix is highly problematic and all
estimators discussed in the electronic supplementary material
[\citet{AKsupp}, Section S.3] yield a poor performance
when testing for changes in the connectome\vadjust{\goodbreak} data set. Therefore, we use
the test statistics $\widetilde{T}_n^{(A/B)}$ as in
(\ref{eqstattilde}) and the bootstrap critical values as described in
Section~\ref{sectionbootstrap} in the analysis of the data set.

\begin{figure}
\begin{tabular}{@{}c@{}}

\includegraphics{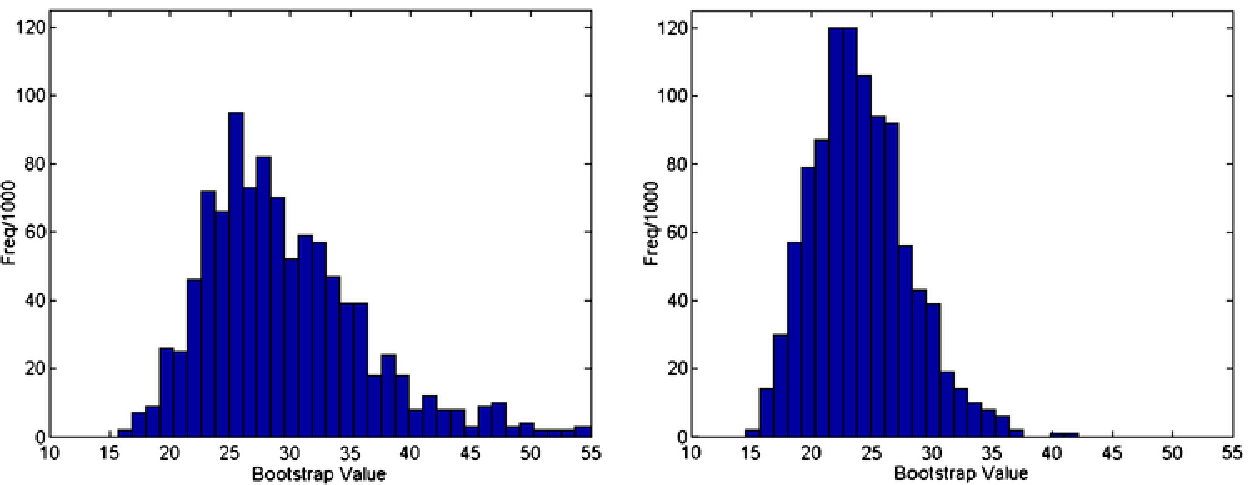}
\\
\textup{(a)}\\[6pt]

\includegraphics{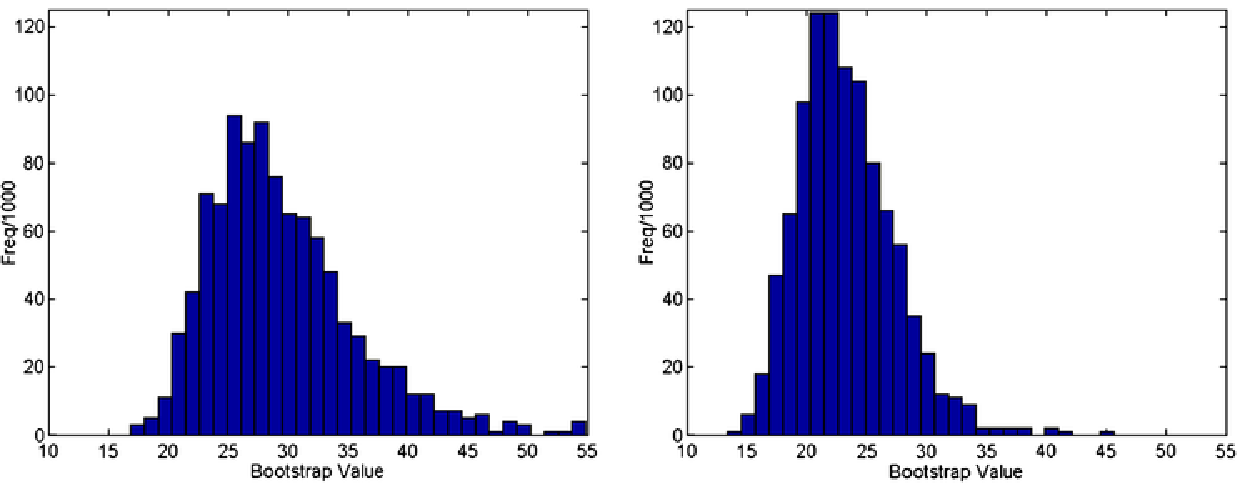}
\\
\textup{(b)}
\end{tabular}
\caption{Bootstrap distributions for four randomly chosen scans,
\textup
{(a)} two
with changes detected, \textup{(b)}~two with no changes detected, when
using 125
components and the sum-statistic $\widetilde{T}_n^{(A)}$. The
distributions vary due to the differing temporal correlation structures
for different individuals.}\label{figbootstrap}
\end{figure}

\begin{figure}

\includegraphics{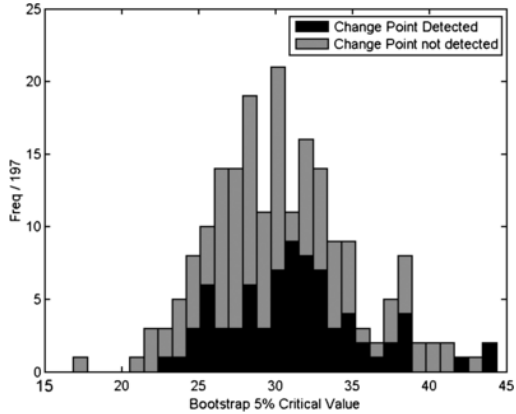}

\caption{Distribution of bootstrap 5\% critical values from 197 scans,
where the stacking shows whether the critical value was from a scan
with detected or no detected change using 125 components and the
sum-statistic $\widetilde{T}_n^{(A)}$.}\label{figrealdataa}
\end{figure}

Figure~\ref{figbootstrap} shows four typical examples of bootstrap
distributions with and without changes detected. While differences due
to the different underlying correlation structures are clearly visible,
no difference is apparent between scans which contain a detected change
and those which do not. Figure~\ref{figrealdataa} shows the
distribution of the 5\% bootstrap critical values from 197 scans, once
more indicating that the critical values show some deviation between
scans due to different underlying correlation structures, hence
different limit distributions, but do not differ between those with or
without changes detected.

After the preprocessing of the data described in Section \ref
{secsinglefmri}, a separable functional principal component
decomposition was found,\vadjust{\goodbreak} based on the three orthogonal directions
within the image acquisition. Eigen-decompositions of the empirical
covariance functions were used to generate the full three-dimensional
\mbox{functional} basis. The eigenvalues associated with the decompositions
did not decrease particularly fast. Indeed, the first 1000 eigenvalues
only explained approximately 5\% of the variation. In many
applications, this is unappealing as it means that the data cannot be
sparsely represented. However, in change-point detection, a~flat
eigenstructure in the uncontaminated covariance can actually (and
somewhat counter-intuitively) enhance detectability and is therefore
actually an advantageous \mbox{property}. By Corollary~\ref{lemseppower},
change-points, if present, will tend to be found in eigenfunctions with
larger relative eigenvalues, and hence only a small number of
\mbox{components} need to be checked, especially when the components are flat.
Thus, the number of components to examine was set to a small number,
namely, systems with 64 ($=4^3$) and 125 ($=5^3$) eigenfunctions were
investigated, with each direction having either its top 4 or 5
eigenfunctions as part of the tensor product. This was a compromise
between having a large number of components, which would reduce the
finite sample detectability as well as computational speed (processing
time in Matlab for one scan with 1000 bootstrap samples for 125
components was approximately 6--7 hours on a desktop PC, while
processing for the entire 197 scans took approximately 24 hours on a 40
node cluster), and having a sufficient number of components not to miss
possible changes. Since the original data set was of dimension
$64\times64\times33$, systems with 64 and 125 eigenfunctions
correspond to an approximate dimension reduction by a factor of 2000 or
1000, respectively. Three examples of the projected data of dimension 64
were discussed in Section~\ref{secmodel}.

The test statistics $\widetilde{T}_n^{(A/B)}$ in (\ref{eqstattilde})
were found for all 197 scans for a change-point. Bootstrap resampling
as described in Section~\ref{sectionbootstrap} was used to obtain
critical values for each time series ($M=1000$). Multiple comparisons
were corrected controlling the FDR by the procedure of \citet
{BenjaminiHochberg} for independent observations. In this case, unlike
in usual brain imaging applications, the correction is done across
subjects, not across space, as here space is a single functional
observation, while different subjects can be deemed independent.

\begin{table}
\caption{Results of the 64 and 125 component analyses. ``No
correction''
indicates all rejections at the 5\% level were counted, while ``FDR
correction'' indicates FDR correction was used at a 5\% level, with the
corresponding threshold being given}\label{tabresults}
\begin{tabular*}{\tablewidth}{@{\extracolsep{\fill}}lcccc@{}}
\hline
\textbf{Number of} &&\textbf{Rejections}&\textbf{Rejections}&\\
\textbf{components}&\textbf{Statistic used}&\textbf{(no correction)}
&\textbf{(FDR correction)}&\textbf{FDR thresh}\\
\hline
\hphantom{0}64 &max($\widetilde{T}_n^{(B)}$)&\hphantom{0}88&\hphantom{0}85&0.025\\
&sum($\widetilde{T}_n^{(A)}$)&\hphantom{0}78&\hphantom{0}70&0.022\\
[6pt]
125 &max($\widetilde{T}_n^{(B)}$)&109&107&0.029\\
&sum($\widetilde{T}_n^{(A)}$)&\hphantom{0}82&\hphantom{0}76&0.022\\
\hline
\end{tabular*}
\end{table}

The test results are summarized in Table~\ref{tabresults}. There was
not a large difference whether 64 or 125 components were chosen,
particularly for the sum statistic. Indeed, a small number of subjects
became insignificant when 125 components instead of 64 components were
used, while others became significant. Therefore, the results look
fairly stable regardless of the number of components chosen. If the sum
statistic is used, approximately 40\% of all subjects in the study were
found to have some form of nonstationarity present, which resulted in
their being rejected as stationary against an epidemic alternative.

\subsection{Comparison of results to exponentially weighted moving
average method}

An alternative method for determining change-points is that given by
Lind\-quist, Waugh and Wager (\citeyear{LindquistWW2007}) where an exponentially
weighted moving average
(EWMA) scheme is adopted. This is based on control chart theory and
uses control limits to determine periods of switching between states.
The method has been shown to be particularly appropriate in tasks where
activations take place, but where the times of onset and duration are
not known.

The methodology has two principle differences from the approach adopted
in this paper. First, it is a voxelwise approach as opposed to a
functional approach. This means that each voxel is tested individually.
While this has the obvious advantage of being able to determine on a
voxel by voxel basis if changes occur,\vadjust{\goodbreak} it has the disadvantage that
multiple comparisons need to be taken into account, and also the times
of changes need not be similar even among neighboring voxels, yielding
difficulties in interpretation. The second major distinction is that
the approach requires a parametric model for the error structure, as
opposed to the nonparametric approach within the method proposed in
this paper. The choice of error structure is known to affect the
detection of change-points if incorrectly specified, and indeed has
been shown to be problematic for fMRI time series in particular [see,
e.g., \citet{Nametal2012}].

\begin{figure}

\includegraphics{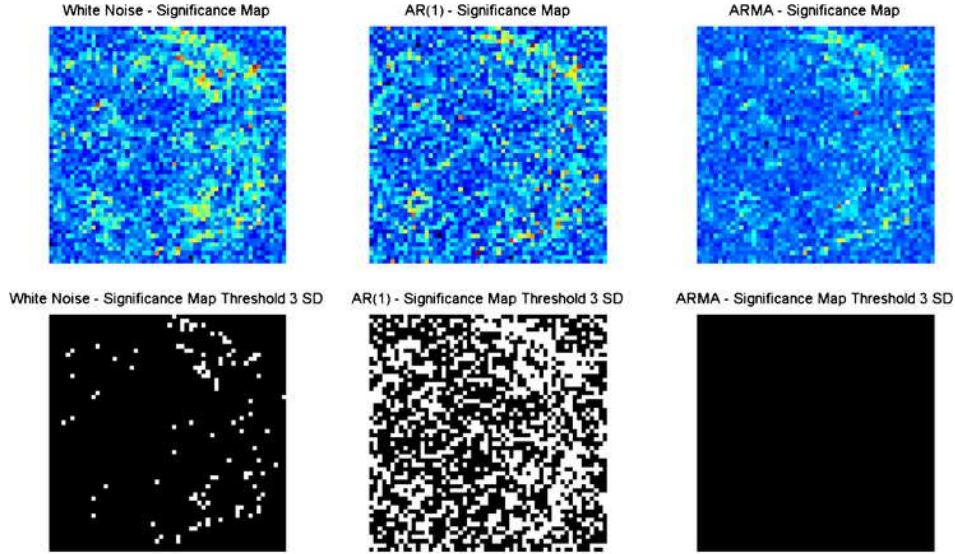}
 \caption{Results for three different noise model
assumptions for the EWMA method as applied to subject 01018 (view of
plane 15). The top row shows the significance maps for white noise,
$\operatorname{AR}(1)$ errors and $\operatorname{ARMA}(1,1)$ errors, respectively, while
the bottom row shows those voxels which would be deemed significant at
a threshold of 3 standard deviations. As can be seen, the results of
the method depend considerably on the noise model
chosen.}\label{figewma}
\end{figure}

Resting state data is inherently different from activation data and the
model for the noise will be inherently more important in this case, in
that no activation is expected to take place. As can be seen in Figure
\ref{figewma}, depending on whether a white noise, AR(1) or ARMA($1,1$)
%
model is chosen, the number of change-points within the image varies
considerably, despite the same threshold being applied. The same
analysis using the methodology proposed in this paper resulted in
nonstationarities being detected (see Figure \ref
{figcomponentseriesstrong}). The differences in the EWMA analysis
for alternative noise models are likely due to the difficulty in
expressing the noise structure accurately for resting state data, in
comparison to activation-baseline tasks where AR(1) and ARMA($1,1$) type
noise structures are known to be fairly good approximations.\vadjust{\goodbreak}

\section{Distribution of the position and
duration of the epidemic change}\label{sectionhierarchical}
The discussion in the previous sections has dealt with situations where
one functional time series is observed and for this time series the
question arises if and when a change has occurred. In some situations,
such as in psychological experiments or in stress testing, due to the
design of the experiment [cf., e.g., \citet{LindquistWW2007}], one can
be reasonably sure that a certain change will occur. Usually in such
situations more than one time series, namely, one time series for each
person involved in the experiment, is observed.
Therefore, it makes sense to include the change-point in the model and
estimate the density of the change-point. For example, one may be
interested in knowing the distribution of the change-point in stress
testing to get an idea about the change and duration distribution.

\subsection{Density estimation of the change-point for hierarchical
time-series}\label{sectiondensity}

Before giving technical details, let us summarize the results of this
subsection as follows. First, it is possible to show that even if we
use the estimated change-point as derived earlier instead of the true
change-point, the empirical distribution function (EDF) and the kernel
density estimate (KDE) of the joint epidemic change-point location and
duration both remain consistent. In the case of fMRI, this allows us to
take the change-points positions from each subject and combine them to
give a population based distribution of the times of changes that occur
in the scanner. By showing that both the EDF and KDE are valid means
that either a histogram based approach or a smooth density approach can
be used as required. As the change-points are functions of time, they
can be combined across subjects without requiring spatial
normalization, because the distributions are independent of the spatial
location of the change. In fact, there may be many different causes of
a nonstationary change in the data, with the question arising as to
whether these might have consistent timings within the scanning period.

In the remainder of the section we give the results for EDFs and KDEs
in full statistical details. Those readers most interested in the
results of such estimates for fMRI resting scan data could proceed to
Section~\ref{sectionfMRI} where the data analysis is detailed.

Let in case of AMOC
\[
X_{t,j}(u)=Y_{t,j}(u)+\mu_{j}(u)+
\Delta_{j}(u) 1_{\{t>\vth_j n\}},\qquad 1\leq t\leq n,1\leq j\leq m,
\]
where the $m$ observed functional time series $\{X_{t,1}\dvtx1\leq t\leq n\}
,\ldots,\{X_{t,m}\dvtx1\leq t\leq n\}$ are independent, $\{ {{\bolds
{\mu
}}_{j}}\dvtx1\leq j\leq m\}$, $\{\bolds{\Delta}_j\dvtx1\leq j\leq m\}$, and
$\{
\vth_j\dvtx1\leq j\leq m\}$ are no longer fixed deterministic but rather
i.i.d. random variables independent of $\{Y_{t,j}(\cdot)\dvtx t\geq1\},
j=1,\ldots,m$, $P(0<\vth_1<1)=1$ and $P(\bolds{\Delta
}_{1}\equiv
0)=0$. For each fixed $j$, the model is still as before, and the index
$j$ indicates the person to whom the observation belongs.

Furthermore, we assume $n=n(m)\to\infty$ as $m\to\infty$.

Denoting $P^*(\cdot)=P(\cdot| \vth_j,{\bolds{\mu
}_{j}},{\bolds{\Delta}_{j}},j=1,\ldots,m)$, the consistency
property $|\widehat{\vth}-\vth|=o_P(1)$ of AMOC estimators [cf. Theorem
2.3 in \citet{AK2}] in the standard setting as outlined in Section
\ref{sectionCPD} translates into
%
\begin{equation}
\label{eqcondconv} |\vth_j-\widehat{\vth}_j|=o_{P^*}(1)\qquad
\mbox{a.s.}
\end{equation}
if the assumptions are a.s. fulfilled, that is, the mean changes are
a.s. nonor\-thogonal to the contaminated projection subspace and the
basis is an orthonormal system almost surely.
%
\begin{theorem}\label{lemdfest}
If (\ref{eqcondconv}) holds and the distribution function $F_{\vth}$
of $\vth$ is continuous, then
\[
\widehat{F}_{\widehat{\vth},m}(x):=\frac1 m \sum_{j=1}^m1_{\{
\widehat
{\vth}_j\leq x\}}
\]
is a consistent estimator for $F_{\vth}$, that is,
\[
\sup_{x\in[0,1]}\bigl\llvert \widehat{F}_{\widehat{\vth},m}(x)-F_{\vth}(x)
\bigr\rrvert \to0 \qquad\mbox{a.s.}
\]
\end{theorem}

The following theorem gives a corresponding result for kernel density
estimators if a rate for the estimators of the change-point
[analogously to (\ref{eqestrate})] is available.
%
\begin{theorem}\label{lemdensest}
Let $h=h(m)\to0, hm\to\infty$ as $m\to\infty$. Assume
%
\begin{equation}
\label{eqcondconvrate} h^{-1}|\vth_j-\widehat{
\vth}_j|=o_{P^*}(1) \qquad\mbox{a.s.},
\end{equation}
which follows, for example, from the analogue of (\ref{eqestrate}) if
$h^2n\to\infty$.
Let $K(\cdot)$ be a bounded and Lipschitz continuous kernel [$K(\cdot
)\geq0$, $\int K(x) \,dx=1$], then
\[
\int\E\bigl\llvert \widehat{f}_{\widehat{\vth},m}(x)-\widehat {f}_m(x)
\bigr\rrvert^2 \,dx\to0,
\]
where
\[
\widehat{f}_{\widehat{\vth},m}(x)=\frac{1}{mh} \sum
_{i=1}^mK \biggl( \frac
{x-\widehat{\vth}_i}{h} \biggr)
\]
and
\[
\widehat{f}_{m}(x)=\frac{1}{mh}\sum
_{i=1}^mK \biggl( \frac{x-{\vth}_i}{h} \biggr)
\]
is the standard kernel estimator of the density $f_{\vth}$ of $\vth$.
\end{theorem}

The theorem shows, in particular, that under standard assumptions on
the kernel and the density it holds
\[
\int\E\bigl\llvert \widehat{f}_{\widehat{\vth},m}(x)-f_{\vth}(x)\bigr
\rrvert^2 \,dx\to0.\vadjust{\goodbreak}
\]

\begin{rem}
For the univariate problem one can show
\[
P \bigl( \llvert \widehat{\vth}-\vth\rrvert \geq c_n \bigr)\leq C
\bigl(\min(\vth,1-\vth)\bigr)^{-2}\Delta^{-2}
n^{-1}c_n^{-1},
\]
where $C$ does not depend on $\vth$ or $\mu,\Delta$; cf., for example,
\citet{kokleip98}. If additionally $\E[\Delta^{-2}\min(\vth_1,1-\vth_1)^{-2}]<\infty$, then using the Markov-inequality and similar
arguments as in the proof of the above theorem, one can conclude
\[
\sup_x\bigl\llvert f_{\widehat{\vth},m}(x)-\widehat{f}_m(x)
\bigr\rrvert \to 0 \qquad\mbox{a.s.},
\]
if, for example, $nh^3, mh^3\to\infty$.
This shows that in this situation under standard assumptions it holds
$\sup_x\llvert f_{\widehat{\vth}}(x)-f_{\vth}(x)\rrvert \to0$ a.s.
\end{rem}

If we are interested in estimators for an epidemic change, things
become slightly more complicated. The above results carry over
immediately to $\widehat{\vth}_i=\widehat{\vth}_{1i}$ as an estimator
for the first change-point as well as to $\widehat{\tau}_i=\widehat
{\vth}_{2i}-\widehat{\vth}_{1i}$ as an estimator for the duration of the
epidemic change, so the marginal distributions can be estimated this
way. This gives the joint distribution if one assumes that the first
change-point ${\vth}_{1i}$ and the duration of the epidemic change
${\tau}_i$ are independent [as, e.g., done by \citet{LindquistWW2007}].
If one does not want to make this assumption, one can formulate an
analogous result using a two-dimensional kernel $K(x,y)$, that is,
$\int K(x,y)\,dx \,dy=1$, that is positive and bounded, and fulfills the
following Lipschitz condition
\[
\bigl|K(x_1,y_1)-K(x_2,y_2)\bigr|\leq C
\max\bigl(|x_1-x_2|,|y_1-y_2|\bigr)
\]
for some $C>0$. Then, if $mh_1h_2\to\infty$, $h_1,h_2\to0$, one gets
an analogous result as in Theorem~\ref{lemdensest} for
\begin{eqnarray*}
\widehat{f}_{\widehat{\vth}_i,\widehat{\tau}_i,m}(x,y)&=&\frac
{1}{mh_1h_2}\sum
_{i=1}^mK \biggl(\frac{x-\widehat{\vth
}_i}{h_1},
\frac
{y-\widehat{\tau}_i}{h_2} \biggr),
\\
\widehat{f}_{m}(x,y)&=&\frac{1}{mh_1h_2}\sum
_{i=1}^mK \biggl(\frac
{x-{\vth
}_i}{h_1},
\frac{y-{\tau}_i}{h_2} \biggr).
\end{eqnarray*}
The proof is analogous to the proof of Theorem~\ref{lemdensest}.

\subsection{Estimation for the connectome resting state data}
\label{sectionfMRI}

The results in the previous section can now be applied for the subjects
that survived the FDR threshold as outlined in Section \ref
{secfMRIboot}, and the joint distribution of position and duration of
the epidemic change can be derived.

\begin{figure}

\includegraphics{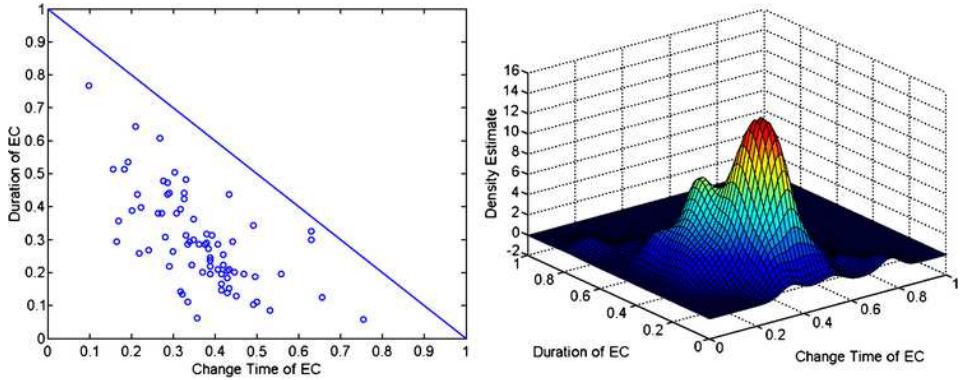}

\caption{Estimators for 76 fMRI scans surviving FDR correction based on
125 components and the sum statistic $\widetilde{T}_n^{(A)}$.
Left: joint estimates of position and duration of epidemic change.
Right: kernel density estimate using a Gaussian kernel and bandwidths
$h_x=0.04, h_y=0.05$.}\label{figrealdata}
\end{figure}

The left panel in Figure~\ref{figrealdata} shows the estimated change
and durations for all those subjects where the null hypothesis of no
change was rejected using FDR, while the right panel shows a kernel
smoothed density estimate for the joint distribution\vadjust{\goodbreak} of position and
duration of the epidemic change, using the automatic bandwidth
selection procedure of \citet{BotevGK2010} (yielding bandwidths of
$h_x=0.04$ and $h_y=0.05$). In this example change-points usually occur
somewhere between 0.25 and 0.5, and last around 0.1--0.3 of the scanning
period except for very early changes which often last longer. In fact,
the density seems to be bimodal, indicating two clusters dividing
subjects into those for which a change occurs after a relatively short
period in the scanner (maybe only now arriving in the stationary state)
in addition to a relatively long duration (possibly until the end of
the scan), and those subjects for which after a short time in the
epidemic state a return to baseline happens. However, for subjects with
a relatively late change, a long duration cannot happen due to the
limited time in the scanner. Therefore, the two modes may be an
artifact of the statistical procedure based on the short time span.

The results of the study show that resting state scans in some cases do
show evidence of deviation from stationarity which can be modeled by
epidemic mean changes, at least as a first approximation, indicating
that the overall activity is different at different times. This result
has implications for studying correlations within the brain between
regions of interest using multiple subjects, particularly if some
subjects show nonstationary behavior, while others do not.

\section{Conclusions}\label{sectionconclusions}

In this paper a methodology for the detection and estimation of
change-points from multiple subjects has been outlined, and the
associated statistical properties investigated. It has been shown that
change-point analysis is a useful tool in situations where very
high-dimensional data sets are collected across time, especially if the
data have a natural spatial structure.\vadjust{\goodbreak} One main result explains the
impact of the choice of projection subspace estimation on the power of
the tests. In particular, any structural breaks present will likely be
found within the first few components when the eigenspectrum is
relatively flat if one uses estimated principal components for the
projection. The second main result shows that consistent estimators for
the change-points exist and the associated distribution of change-point
locations and durations
can be found.

The main aim of this paper was to find a general framework for the
testing and estimation of change-points in resting state fMRI data, in
such a way that details such as the estimation procedure for the
projection subspace can be replaced with different statistical
techniques while the underlying theoretical results remain valid.
Examples include methodology based on fixed spatial basis choices such
as wavelets, or computational methods such as those by \citet
{Zipunnikovetal2010} extended to time series settings. For these
variations, by careful choice of the estimators for the projection
subspace, tests as well as estimators for the location and duration
distributions can be obtained from the theoretic results given in this paper.

For resting state fMRI data, the covariance function $c$ is probably
one of the most important quantities of interest. Indeed, the full
function would give a complete connectivity map for the brain. However,
due to its inherent size, connectivity studies take approximations or
subsets of this function and use these to derive models for the default
network, for example. However, as we have seen in the theoretical
analysis, when nonstationarities are present, we do not observe $c$ but
rather the contaminated version $k$, that is, the connectivity map but
also elements associated with nonstationarities. As there is no
inherent reason to believe these nonstationarities are anything other
than subject specific, they will induce false correlations not related
to the true underlying connectivity in a standard correlation based
analysis to derive connectivity measures. However, by performing tests
such as those we have derived, it is at least now possible to pick
candidate subjects with no evidence of nonstationarities, or
alternatively investigate further the causes of the nonstationarities
in those with evidence of such changes, in case they are intrinsically
part of the default network in multiple subjects.

It should be noted that while we have used tests and estimators
designed for epidemic changes in this paper, it is likely that other
forms of nonstationarity might be present in applications such as fMRI,
as well as possible multiple epidemic changes. However, the use of
epidemic changes is a good first approximation as it not only mimics
the most likely form of nonstationarity present in fMRI but will also
have power against other alternatives too, including multiple epidemic
changes as well as slow transient changes (where instead of a jump up
or jump back, this takes some amount of time). Of course, the detection
will not be optimal in these cases, but detection is still likely for
reasonable sized changes. It is for this reason that we feel that it
would be unwise to draw too many conclusions from the actual maps that
could be\vadjust{\goodbreak} generated for $\Delta(u)$ based on the epidemic change
alternative. However, because of this, neuroscientific conclusions
should really be restricted to those which can be based on the timings
of changes, with further investigation being required, to account for
possible effects such as hemodynamic lag, to draw conclusions
concerning any underlying neuronal changes.

While the estimators and tests can be used in many applications, from
epidemics to image based security surveillance, the application that
drove all the theoretical developments was resting state fMRI. As a
result, for future statistical analyses of resting state fMRI data,
this study has three main implications:
\begin{itemize}
\item First, routine testing for nonstationarities in resting-state
scans is now possible, and relatively computationally inexpensive
(compared to the time taken to do further analyses).
\item Second, this study indicates that the examined subjects are
fairly well split between those that have evidence of nonstationarities
and those who do not, so that it would be of great interest to compare
the connectivity relationships between these two groups. Many of the
most standard connectivity measures are based on correlation analyses,
which can be dramatically affected by the presence of
nonstationarities. Hence, investigation of the phenomena found in this
paper warrants further exploration.
\item Third, the distributions derived from the change-point estimators
seem to indicate that the location and duration of the
nonstationarities have considerable mass around half way through the
scan. This position (in contrast to the test result) could be a
statistical artifact, in that while the test itself reveals the
presence of nonstationarity, the type of nonstationarity might not be
epidemic, but the epidemic change hypothesis could still be powerful
against evidence of stationarity. It would thus be of interest to
investigate further whether this nonstationary behavior is due to the
ability or inability to rest within the scanner and is due to active
thought processes interrupting the resting state network, or whether
the resting state signal itself changes after a certain amount of time.
This could be investigated by looking at the spatial distribution of
the time series which exhibit changes, but requires further statistical
development to rigorously allow the examination of individual spatial
maps after the omnibus test for the presence of an epidemic change.
\end{itemize}

\section*{Acknowledgment}

John A. D. Aston thanks SAMSI for hosting the author during which some
of the work was carried out.

\begin{supplement}
\stitle{Supplementary material for evaluating stationarity via
change-point alternatives with applications to fMRI data}
\slink[doi]{10.1214/12-AOAS565SUPP} 
\slink[url]{http://lib.stat.cmu.edu/aoas/565/supplement.pdf}
\sdatatype{.pdf}
\sdescription{The supplementary material provides added technical
details along with the proofs of the results in the paper.}
\end{supplement}


\printaddresses

\end{document}